\documentclass[aps,pre,floatfix,twocolumn]{revtex4}
\usepackage{epsf,colordvi,color,graphicx,epsfig}
\usepackage{amsmath,wasysym,color}
\usepackage{ulem}
\usepackage[arrow, matrix, curve]{xy}
\usepackage{subfigure}
\usepackage[T1]{fontenc}

\begin{document}

\title{Boolean versus continuous dynamics on simple two-gene modules}
\author{Eva Gehrmann and Barbara Drossel}
\address{Institut f\"ur Festk\"orperphysik, TU Darmstadt, Hochschulstra\ss e 6, 64289 Darmstadt, Germany}

\begin{abstract}
 We investigate the dynamical behavior of simple modules composed of two genes with two or three regulating connections. Continuous
dynamics for mRNA and protein concentrations is compared to a Boolean model for gene activity. Using a generalized method, we study within a single framework different continuous models and different types of regulatory functions, and establish conditions under which the system can display stable oscillations. These conditions concern the time scales, the degree of cooperativity of the regulating interactions, and the signs of the interactions. Not all models that show oscillations under Boolean dynamics can have oscillations under continuous dynamics, and vice versa. 
\end{abstract}

\maketitle

\section{Introduction}
\label{introduction}
Soon after the first repressor protein had been discovered by Fran\k{c}ois Jacob and Jacques Lucien Monod in 1961~\cite{jacob1961}, theoretical work on gene regulation started \cite{monod1963}. Typically, a realistic modelling of gene regulation systems includes rate equations for the concentrations of the participating macromolecules, i.e., mRNA and proteins.  In vivo experiments permitted to obtain quantitative data for regulatory processes and their kinetic parameters, and they provided important insight into regulatory dynamics. Thus, Elowitz and Leibler~\cite{elowitz2000} designed and constructed a synthetic network out of three transcriptional repressor systems to build an oscillating network, termed the repressilator, in Escherichia coli. In recent years, the field of systems biology has emerged, which aims
at a quantitative description of cell behavior and basic dynamic processes, and which permits to analyze systems such as gene
regulation networks. 

Besides detailed quantitative approaches, the general features of regulatory and signaling processes in living cells also gave rise to a minimalist dynamical description as Boolean networks, where the state of each gene is either ``on'' or ``off''~\cite{kauffman1969,thomas1973}. Such a description is particularly useful when dealing with large networks \cite{bornholdt2005}, because it reduces the huge complexity of the continuous system with its many differential equations and parameters to a problem of logical structure which is easier to understand. It permits to study generic features of entire classes of systems \cite{kauffman1969}, or to reproduce the correct sequence of events in gene regulation networks that must function reliably, such as cell cycle dynamics \cite{li2004}.

So far, little is known about the general conditions under which a Boolean simplification gives a realistic picture of the dynamics in gene regulation networks. In contrast to Boolean networks, ordinary differential equations (ODEs), which model the switch-like dynamics of genes by using sigmoidal Hill functions, can include more detailed information about transcription and translation processes to evaluate the time course of the gene expression patterns. Depending on the parameter values, such models can show oscillating behavior or a stable fixed point. Even for small systems of only two genes, the seems to be no simple relation between Boolean and continuous models. Widder et al.~\cite{widder2007} and Polynikis et al.~\cite{poly2009} studied a two-gene activator-inhibitor network and investigated in detail the 
conditions under which a Hopf bifurcation occurs, which leads to oscillating behavior. They found that oscillatory behavior is exhibited by the two-gene model only if the Hill coefficents are above a certain threshold, and that the system can be driven through the bifurcation by varying the time scales of the mRNA and proteins~\cite{poly2009}. The Boolean version of this system always shows oscillatory behavior.

Similar results can be found in Del Vecchio~\cite{vecchio2007}, who included an additional self-input of one gene and varied the time-scale difference between the activator and the respressor and time-scale difference between the protein and mRNA dynamics by using bifurcation analysis. They also obtained richer dynamical behavior incorporating mRNA dynamics and could define a parameter space that leads to stable limit cycles.

In this paper, we present a general and comprehensive investigation of the two-gene network. Gross and Feudel~\cite{gross2006} developed a 
method of \textit{generalized models}, which allows to investigate the stability of fixed points and the occurrence of bifurcations in 
dependence of general features of the system, without the need to specify the steady state or the regulatory functions. This approach enables us to unite all previous studies of this system within one framework, and to include also those situations that had not been studied before. This permits us to identify the main differences between the dynamical behavior and the attractor patterns of Boolean and continuous models.

The paper is structured as follows. In Section~\ref{model}, we introduce the dynamical equations and the generalized method used for the analysis. Section~\ref{results} presents the conditions for the occurrence of oscillations in the continuous model and compares these to the Boolean model. Finally, we discuss and compare our findings to previous studies in Section~\ref{discussion}.

\section{Model}

\label{model}

Gene expression is the process by which genetic information is ultimately transformed into working proteins. The main steps are transcription from DNA to mRNA, translation from mRNA to linear amino acid sequences and folding of these into functional proteins~\cite{sontag2007}. A certain class of proteins, called transcription factors, can bind to the DNA to regulate the rate at which their target genes are transcribed into mRNA. Gene regulation thus involves a network of macromolecules that mutually influence each other. The production of proteins and mRNA is balanced by degradation and dilution~\cite{chen1999}.\\
\begin{figure}[!ht]
	\begin{center}
		\begin{tabular}{lrr}
			$ \begin{xy} \xymatrix{ 
				a \ar@(ul,dl)[] \ar@/^/[r] & b \ar@/^/[l]
			}
			\end{xy}$
			& ~~~~~~~~~~~~
			$\begin{xy} \xymatrix{ 
				\mbox{mRNA}~a \ar@{.>}[r] 	& \mbox{protein}~a \ar[d] \ar@/_1pc/[l]  \\
				\ar[u] \mbox{protein}~b 	&\ar@{.>}[l]  \mbox{mRNA}~b
			}
			\end{xy}$ 
			\\
\end{tabular}
\caption{An example of a simple gene regulatory network consisting of two genes~$a$ and~$b$ regulating each other, and an additional self-input of gene~$a$. The network consists of four molecular species: mRNA~$a$ and~$b$, and proteins~$a$ and~$b$. Dotted lines label translation processes, continuous lines transcription. Protein~$a$ regulates the production of mRNA~$b$ via activation or inhibition. Protein~$a$ and~$b$ together regulate the production of mRNA~$a$ via a two-dimensional input function.}
\label{bsp}
\end{center}
\end{figure}
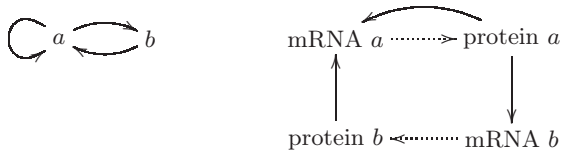

In this paper, we consider a simple gene regulatory network, consisting of two genes~$a$ and~$b$ regulating each other and an additional self-input of gene~$a$. The concentration of mRNA produced by gene~$i$ is denoted by $R_i$, while the corresponding protein concentration is denoted by $P_i$, for $i=a,b$. Translation of mRNA and degradation of mRNA and proteins are supposed to occur with constant rate. Therefore, the ordinary differential equation system describing the reaction kinetics is~\cite{poly2009}

\begin{equation}
	\begin{split}
		\dot{R}_a &= m_a F_1(P_a, P_b) - \gamma_a R_a \\
		\dot{R}_b &= m_b F_2(P_a) - \gamma_b R_b \\
		\dot{P}_a &= \omega_a R_a - \delta_a P_a \\
		\dot{P}_b &= \omega_b R_b - \delta_b P_b 
	\end{split}
\label{ODEs}
\end{equation}

$F_1$ and $F_2$ depend on the concentrations of the regulatory proteins. Regulation by only one protein (such as regulation of mRNA $b$ by
protein $a$) is usually modelled by a monotonically increasing sigmoidal-shaped function when the protein is an activator, and by a decreasing function when the protein is an inhibitor. Experimental evidence~\cite{yagil1971} suggests the usage of \textit{Hill functions}~\cite{hill1910}. Thus, the function $F_2(P_a)$ is either the Hill function for activation

\begin{equation}
F^+(P_a, k_a, n_a) = \frac{P_a^{n_a}}{P_a^{n_a} + k_a^{n_a}}\, ,
\label{activator}
\end{equation}

or the Hill function for inhibition,

\begin{equation}
F^-(P_a, k_a, n_a) = 1 - F^+(P_a, k_a, n_a) = \frac{k_a^{n_a}}{k_a^{n_a} + P_a^{n_a}}\, .
\label{repressor}
\end{equation}

$k_a$ is the \textit{activation coefficient} or \textit{expression threshold} that defines the concentration of protein~$a$ needed to
significantly activate expression. The parameter~$n_a$, called \textit{Hill coefficient}, controls the steepness of the Hill function. The larger~$n_a$, the more step-like is the regulatory function. Biologically, $n_a$ is related to the molecular binding mechanism: it is the number of proteins required for saturation of binding to the DNA~\cite{widder2007} and is therefore also called \textit{cooperativity coefficient}. The Hill function can be considered to be the probability that the promoter region is bound, averaged over many binding and unbinding events of proteins~$i$~\cite{alon}.

\begin{table}
\centering{
\begin{tabular}{ll}
\hline
$R_i$ 		& Concentration of transcribed mRNAs\\
$P_i$		& Concentration of translated proteins\\
$m_i$		& Maximal transcription rates\\
$\omega_i$	& Translation rates\\
$\gamma_i$	& mRNA degradation rates\\
$\delta_i$	& Protein degradation rates\\
$k_i$		& Expression threshholds\\
$n_i$		& Hill coefficients\\
\hline
\end{tabular}
\caption{Notation}}
\label{notation} 
\end{table}

\begin{figure}[!ht]
\centering{
\includegraphics[width=0.45\textwidth]{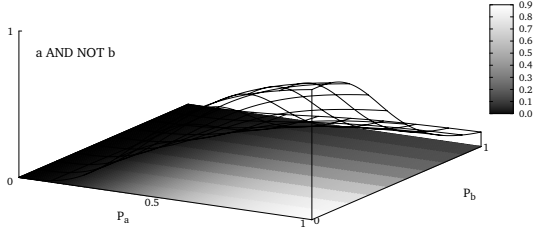}
\caption{HillCube of the input-function $F=P_a$~AND~NOT~$P_b$. The concentration of transcribed mRNA is projected to the two-dimensional surface. In light-colored regions, the concentration of activator,~$P_a$, is high, and the concentration of inhibitor,~$P_b$, is low, so mRNA~$a$ is transcribed significantly.}
\label{hillcubes}
}
\end{figure}

The combined effect of multiple transcription factors is described by using multi-dimensional input functions.  An example for deriving such
a function from gene expression measurements was given by Setty~et~al.~\cite{setty2003}, who used the \textit{lacZYA} operon of Escherichia coli. The function $F_2(P_a,P_b)$ can, for instance, integrate an activator~$P_a$ and a repressor~$P_b$~\cite{alon}. If the activator and the inhibitor bind to the promotor independently, there are four binding-states of the promotor: unbound, bound to either $P_a$ or $P_b$, or bound to both proteins. Transcription occurs mainly in the case that the activator binds the promoter and the repressor does not, resulting in a $P_a$~AND NOT~$P_b$ input function. This is one of the 16 possible Boolean functions for two inputs, and the simplest gene regulation models use  such Boolean functions. The 16 Boolean functions can be represented in $P_a$-$P_b$-space as so-called BooleCubes. Whereever we need to specify the functional form of $F_2(P_a,P_b)$, we will use a continuous generalization of BooleCubes to so-called HillCubes, which is based on Hill functions and was suggested by Wittmann et al.~\cite{wittmann2009}. Thus, the HillCube of the input-function $F_2(P_a,P_b)=P_a$~AND~NOT~$P_b$ is given by the expression $F_2(P_a,P_b)=F^+(P_a,k_a,n_a) \cdot F^-(P_b,k_b,n_b)$ (shown in Fig.~\ref{hillcubes}). The other HillCubes are constructed in analogous manner by taking sums over the appropriate products of Hill functions. For instance, the XOR function is written as $F_2(P_a,P_b)=F^+(P_a,k_a,n_a) \cdot F^-(P_b,k_b,n_b)+F^-(P_a,k_a,n_a)\cdot F^+(P_b,k_b,n_b) $.

In order to investigate the conditions under which model~(\ref{ODEs}) can display oscillations, we will perform a linear stability analysis
of the fixed points, and determine whether a Hopf bifurcation can occur. The fixed point condition is $\dot{R}_a = \dot{R}_b = \dot{P}_a = \dot{P}_b = 0$, leading to ${R_a}^* = \frac{\delta_a}{\omega_a}{P_a}^*$ and ${R_b}^* = \frac{\delta_b}{\omega_b}{P_b}^*$, and to  the implicit equations 

\begin{eqnarray}
P_a^*&=& \frac{\gamma_a\delta_a}{\omega_am_a}F_1(P_a^*,P_b^*)\,\nonumber \\
P_b^*&=&\frac{\gamma_b\delta_b}{\omega_bm_b}F_2(P_a^*)\, .\label{fpimplicit}
\end{eqnarray}

The Jacobian $J$ at this fixed point can be written as

\begin{eqnarray}
J &=& \begin{pmatrix}  \begin{pmatrix}   -\gamma_a & 0 \\ 0 & -\gamma_b  \end{pmatrix} & \begin{pmatrix}   \frac{m_a\partial F_1}{\partial P_a} & \frac{m_a\partial F_1}{\partial P_b} \\ \frac{m_b\partial F_2}{\partial P_a} & 0  \end{pmatrix} \\ 
\begin{pmatrix}   \omega_a & 0 \\ 0 & \omega_b \end{pmatrix}  & \begin{pmatrix}  -\delta_a & 0 \\  0 & -\delta_b \end{pmatrix} \end{pmatrix}\nonumber\\
 &=& \begin{pmatrix} \mbox{mRNA degradation} & \mbox{transcription} \\ \mbox{translation} & \mbox{protein degradation} \end{pmatrix}
\end{eqnarray}

Based on the eigenvalues of the Jacobian, one can determine parameter values at which bifurcations occur. The presence of a single zero eigenvalue indicates the emerge or destruction of fixed points or the exchange of stability properties between two fixed points. Bifurcations of this type are the saddle node bifurcation, the transcritical bifurcation, and the pitchfork bifurcation. This type of bifurcations can be identified by the fact that the determinant of the Jacobian vanishes at the bifurcation point. A second type of local codimension-1 bifurcations is the Hopf bifurcation, characterized by the presence of two purely imaginary eigenvalues of the Jacobian.  If the system is at a stable fixed point before the bifurcation, the steady-state becomes unstable at the bifurcation, and if the bifurcation is supercricital, a limit cycle emerges in the vicinity of the unstable fixed point.  Thus, a Hopf bifurcation implies a transition from stationary to periodic motion. A method that yields analytical test functions for the Hopf bifurcation is the method of resultants (described in detail in~\cite{gross2004}).

Since we intend to investigate model~(\ref{ODEs}) for various types of functions $F_1(P_a,P_b)$ and of parameter values, we use the 
description in terms of \textit{generalized models}, as introduced by Gross and Feudel~\cite{gross2006}.  This method allows to investigate 
the occurrence of different dynamical regions without the need to specify the precise form of interaction terms and equilibrium concentrations. This method has already been successfully applied to ecological food webs~\cite{gross_thesis,gross2006,gross2009}, socio-economic models~\cite{gross2006} and metabolic networks~\cite{steuer2006,steuer2007b}.

With the assumption that there exists at least one fixed point with positive concentrations ${R_a}^*,{R_b}^*,{P_a}^*,{P_b}^*$, one can define normalized state variables $r_i = \frac{R_i}{{R_i}^*}, p_i = \frac{P_i}{{P_i}^*}$ and normalized functions $\tilde{f_j}(p_i)= \frac{F_j({P_i}^* p_i)}{F_j*}$, with the abbreviated notation $F_j^*= F_j(P_i^*)$. The fixed point values then simply are $r_a^*=r_b^*=p_a^*=p_b^*=1$. Rewriting equations~(\ref{ODEs}) in terms of the normalized variables, we obtain 
\begin{equation}
	\begin{split}
		\dot{r}_a &= \frac{m_aF_1^*}{R_a^*} \tilde{f_1}(p_a, p_b) - \gamma_a r_a \\
		\dot{r}_b &= \frac{m_bF_2^*}{R_b^*} \tilde{f_2}(p_a) - \gamma_b r_b \\
		\dot{p}_a &= \frac{R_a^*}{P_a^*}\omega_a r_a - \delta_a p_a \\
		\dot{p}_b &= \frac{R_b^*}{P_b^*}\omega_b r_b - \delta_b p_b.
	\end{split}
\label{ODEs_gen}
\end{equation}

Introducing the four parameters

\begin{equation}
	\begin{split}
		\alpha_r 	&= m_aF_1^*/R_a^* = \gamma_a, \\
		\beta_r 	&= m_bF_2^*/R_b^* = \gamma_a, \\
		\alpha_p 	&= R_a^*\omega_a/ P_a^*= \delta_a, \\
		\beta_p 	&= R_b^*\omega_b/P_b^* = \gamma_b, 
	\end{split}
\label{scale}
\end{equation}

we can rewrite our model as

\begin{equation}
	\begin{split}
 \dot{r}_a &= \alpha_r (\tilde{f_1}(p_a, p_b) - r_a) \\
 \dot{r}_b &= \beta_r (\tilde{f_2}(p_a) - r_b) \\
\dot{p}_a &= \alpha_p (r_a - p_a) \\
 \dot{p}_b &= \beta_p (r_b - p_b).
	\end{split}
\label{ODEs_gen2}
\end{equation}

The Jacobian at a fixed point of this set of equations is

\begin{equation}
 \tilde{J} = \begin{pmatrix} \alpha_r &  &  &  \\  & \beta_r &  &  \\  &  & \alpha_p &  \\  &  &  & \beta_p \end{pmatrix}  \begin{pmatrix} \begin{pmatrix}   -1~ & ~~0~ \\ ~~0~ & -1~ \end{pmatrix} & \begin{pmatrix}   \frac{\partial \tilde{f_1}}{\partial p_a} & \frac{\partial \tilde{f_1}}{\partial p_b} \\ \frac{\partial \tilde{f_2}}{\partial p_a} & ~~0~  \end{pmatrix} \\ 
\begin{pmatrix}   ~~1~~ & ~~0~~ \\ ~~0~~ & ~~1~~ \end{pmatrix}  & \begin{pmatrix}  -1~ & ~~0~ \\ ~~0~ & -1~ \end{pmatrix} \end{pmatrix}
\end{equation}

Throughout this paper, the terms $\frac{\partial\tilde{f_j}}{\partial p_i}$ with $j\in\{1,2\}$ and $i\in\{a,b\}$ will be referred to as new variables $\tilde{f_j}p_i$, which are called \textit{exponent parameters}~\cite{gross2006}. Positive values of $\tilde{f_j}p_i$ indicate that protein~$i$ is considered to be an activator. Negative values indicate inhibition, and if $f_jp_i$ vanishes, protein~$i$ has no influence on the regulatory function $F_j$. Exponent parameters measure the degree of nonlinearity of a process in the steady-state~\cite{gross_thesis}. We can interpret $\tilde{f_2}p_a$ as an indicator of the level of the binding cooperativity of the protein, and therefore of the cooperativity coefficient $n_a$. Using an activating Hill function, we obtain 

$$\tilde{f}(p_a) = \frac{F^+({P_a}^*p_a)}{F^+(P_a^*)} = p_a^{n_a} \frac{P_a^{*n_a} + k_a^{n_a}}{P_a^{*n_a}p_a^{n_a} + k_a^{n_a}}$$
and
$$\left. \frac{\partial \tilde{f}(p_a)}{\partial p_a} \right|_{p_a^*} = \frac{n_a}{1 + (P_a^*/k_a)^{n_a}}\, .$$

Therefore, the exponent parameters $\tilde{f_j}p_i$ are restricted to the interval $[0,n]$, if protein~$i$ is an activator. In the same way, we find that $\tilde{f_j}p_i\in [-n,0]$ if protein~$i$ is a repressor.

Apart from the exponent parameters, the Jacobian also depends on the \textit{scale parameters}~\cite{gross2006} $\alpha_r$, $\alpha_p$, $\beta_r$, and $\beta_p$, which are a measure of the (inverse) time scales of the four concentrations. We assume identical ratios $r=\frac{\alpha_r}{\alpha_p}=\frac{\beta_r}{\beta_p}$ between the mRNA and protein dynamics for both genes.  A large value of $r$ means a quasi steady-state assumption for the mRNA, where the mRNA transients die out quickly.

In the following, we will investigate the properties of the model in dependence of $r$ and the three exponent parameters $\tilde{f_j}p_i$. 

\section{Results}
\label{results}

\subsection{Model with only two connections}

We first consider the case that $F_1$ is independent of $P_a$, leading to the following simplifed network 
\begin{figure}[!ht]
	\begin{center}
		\begin{tabular}{lrr}
			$ \begin{xy} \xymatrix{ 
				a  \ar@/^/[r] & b \ar@/^/[l]
			}
			\end{xy}$
\end{tabular}
\end{center}
\end{figure}

The regulatory functions $F_1$ and $F_2$ can either have the same sign (activator-activator or inhibitor-inhibitor) or different signs (activator-inhibitor complex). In a Boolean model, an activating interaction is implemented by the Boolean function ``copy'', and a repressing interaction by ``invert''. In the Boolean version, the model has only four possible states, 00, 01, 10, and 11. For the activator-activator or inhibitor-inhibitor case, the Boolean model has two fixed points and a cycle. The cycle alternates between 01 and 10 for the activator-activator, and between 00 and 11 for the inhibitor-inhibitor case. For the activator-inhibitor situation, the Boolean  model has a cycle that involves all four states: $00 \to 10\to 11\to 01 \to 00$. We will see that the continuous model can display a cycle only in the activator-inhibitor situation, and only if the parameters are in the appropriate range.

\begin{figure}[!ht]
\subfigure[$r = 1$]{\includegraphics[width=0.13\textwidth]{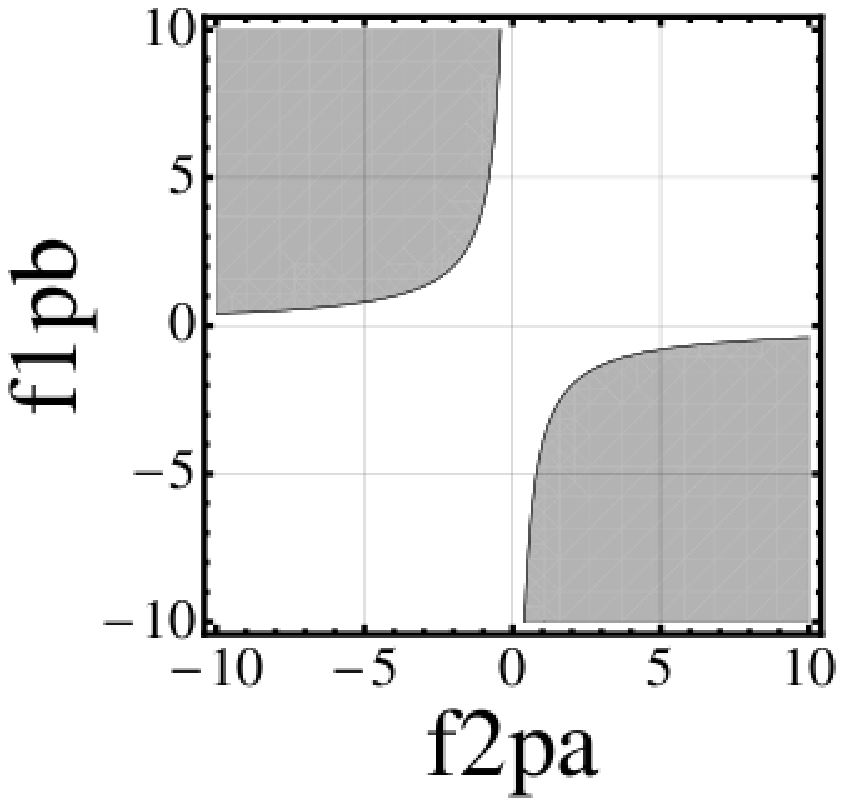}}
\subfigure[$r = 10$]{\includegraphics[width=0.13\textwidth]{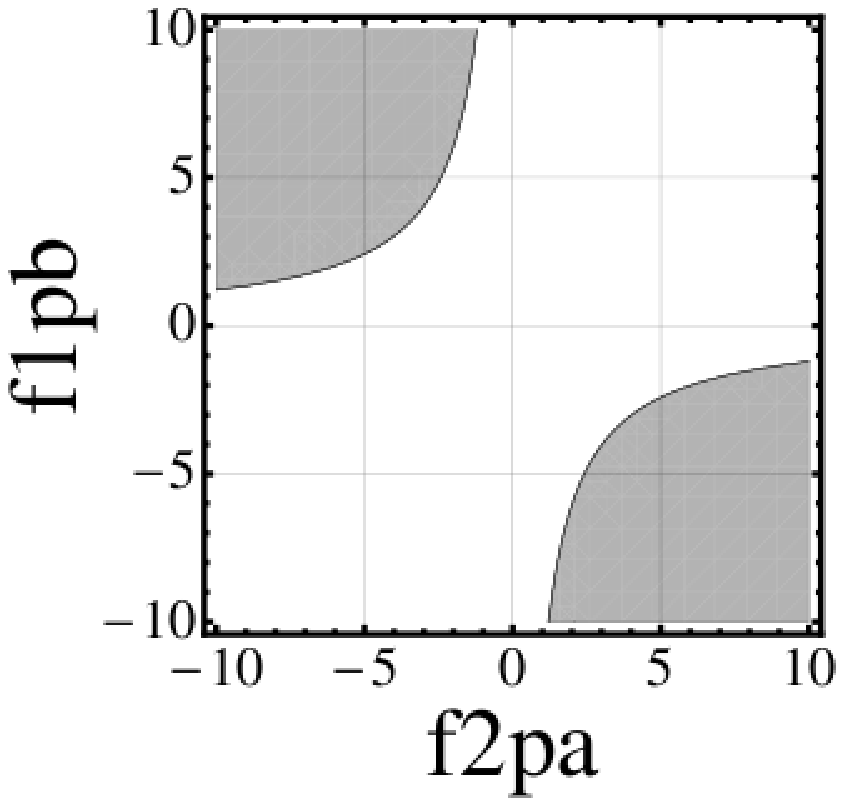}}
\subfigure[$r = 50$]{\includegraphics[width=0.13\textwidth]{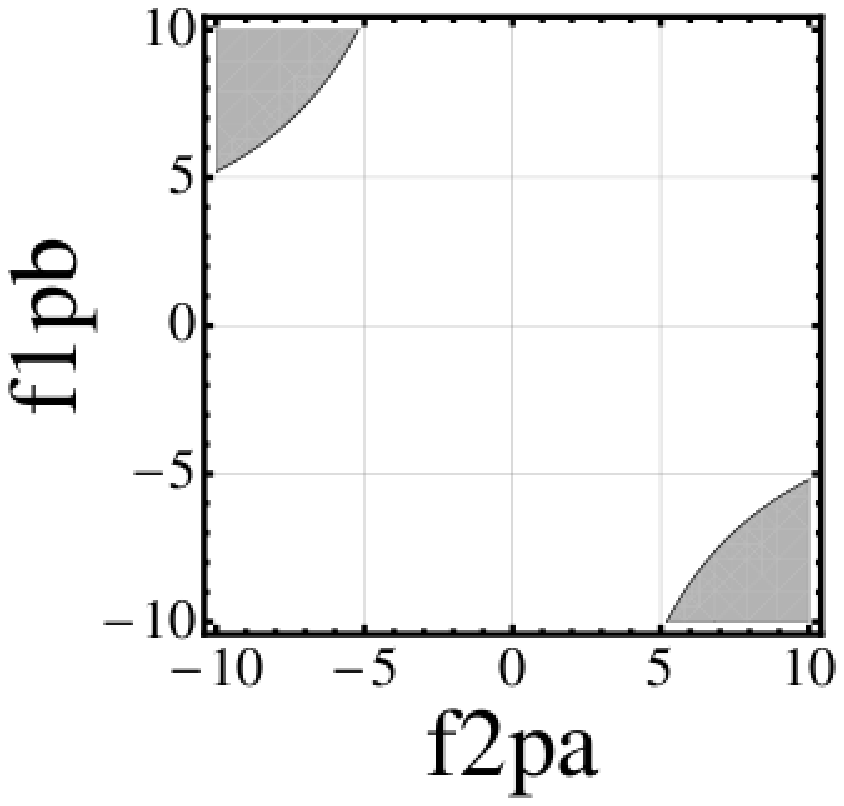}}
\caption{Values of the exponent parameters $\tilde{f_1}p_b$ and $\tilde{f_2}p_a \in\{-10,10\}$, for which the fixed point is unstable and has a complex conjugate pair of eigenvalues with a positive real part.  The time scale ratio $r$ between mRNA and protein dynamics increases from left to right.}
\label{varied_r_2D}
\end{figure}

\begin{figure}[!ht]
\centering
\subfigure{\includegraphics[width=0.23\textwidth]{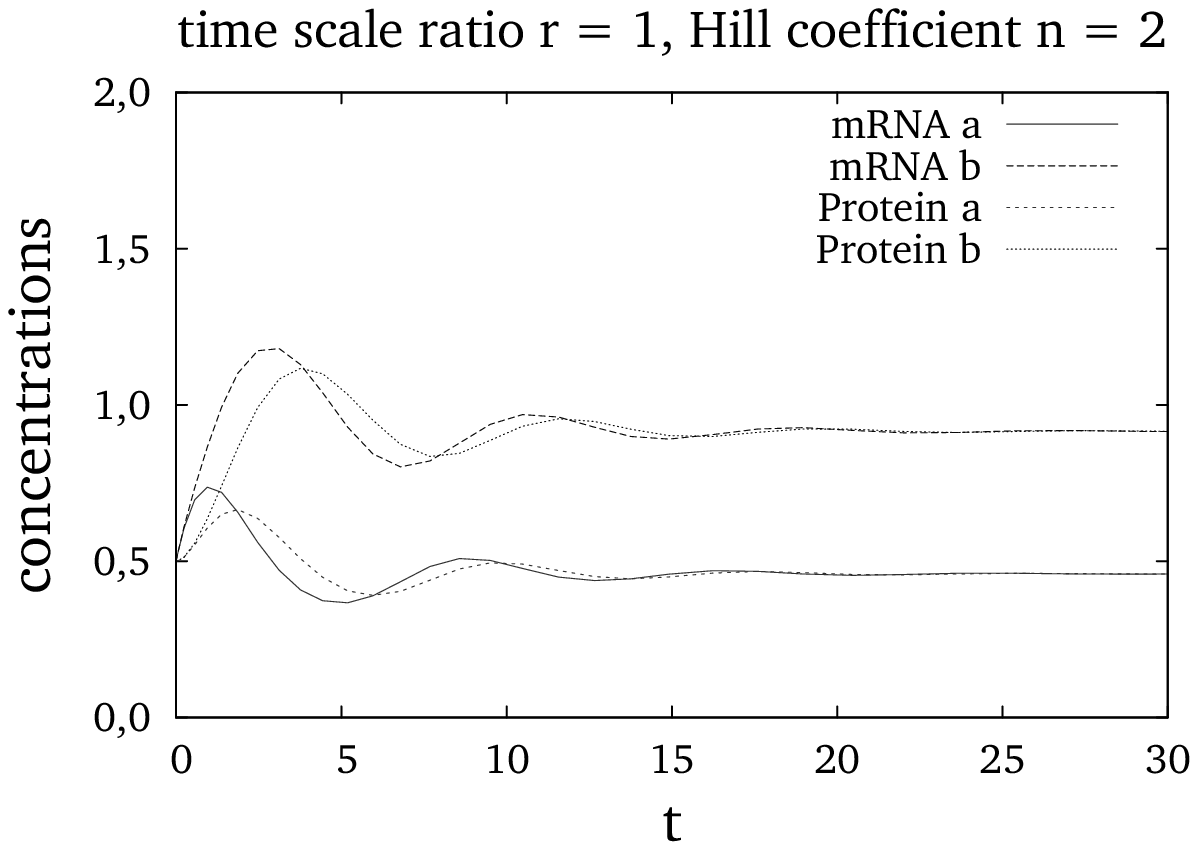}}
\subfigure{\includegraphics[width=0.23\textwidth]{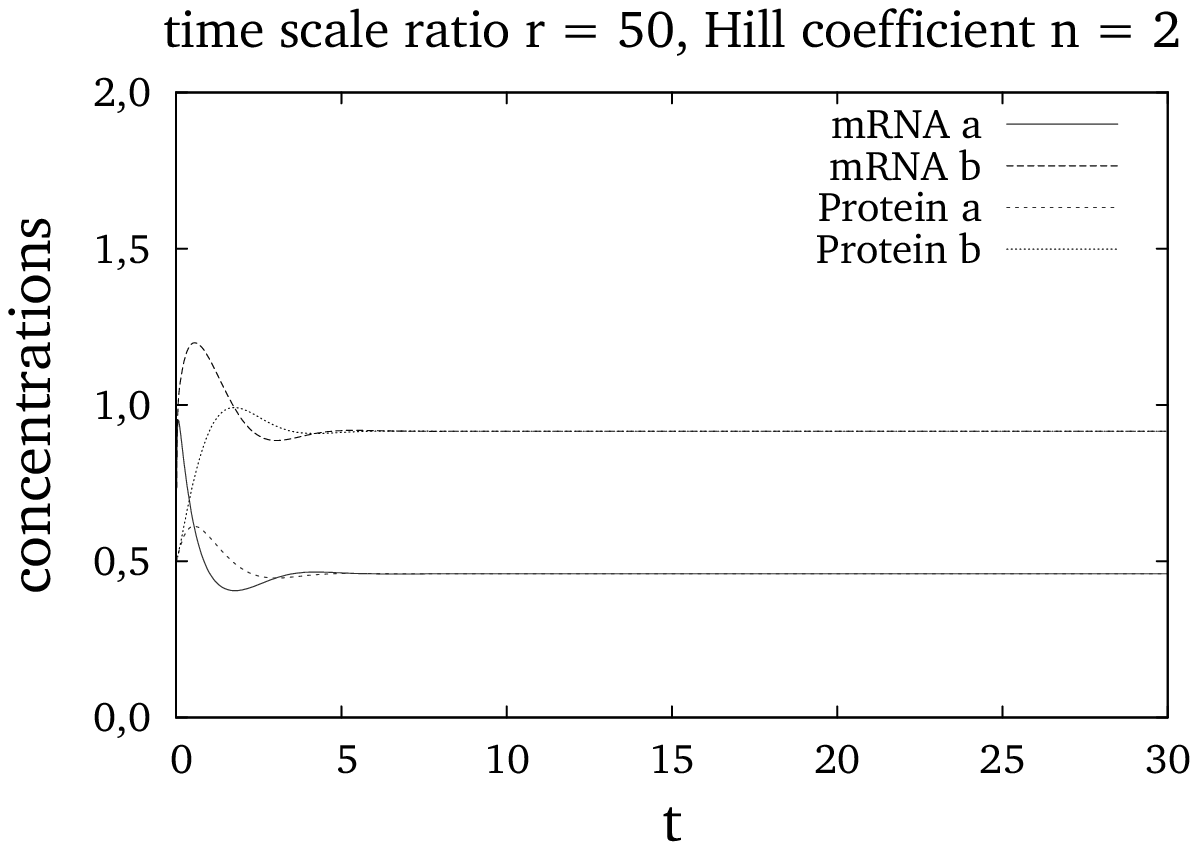}}\\
\subfigure{\includegraphics[width=0.23\textwidth]{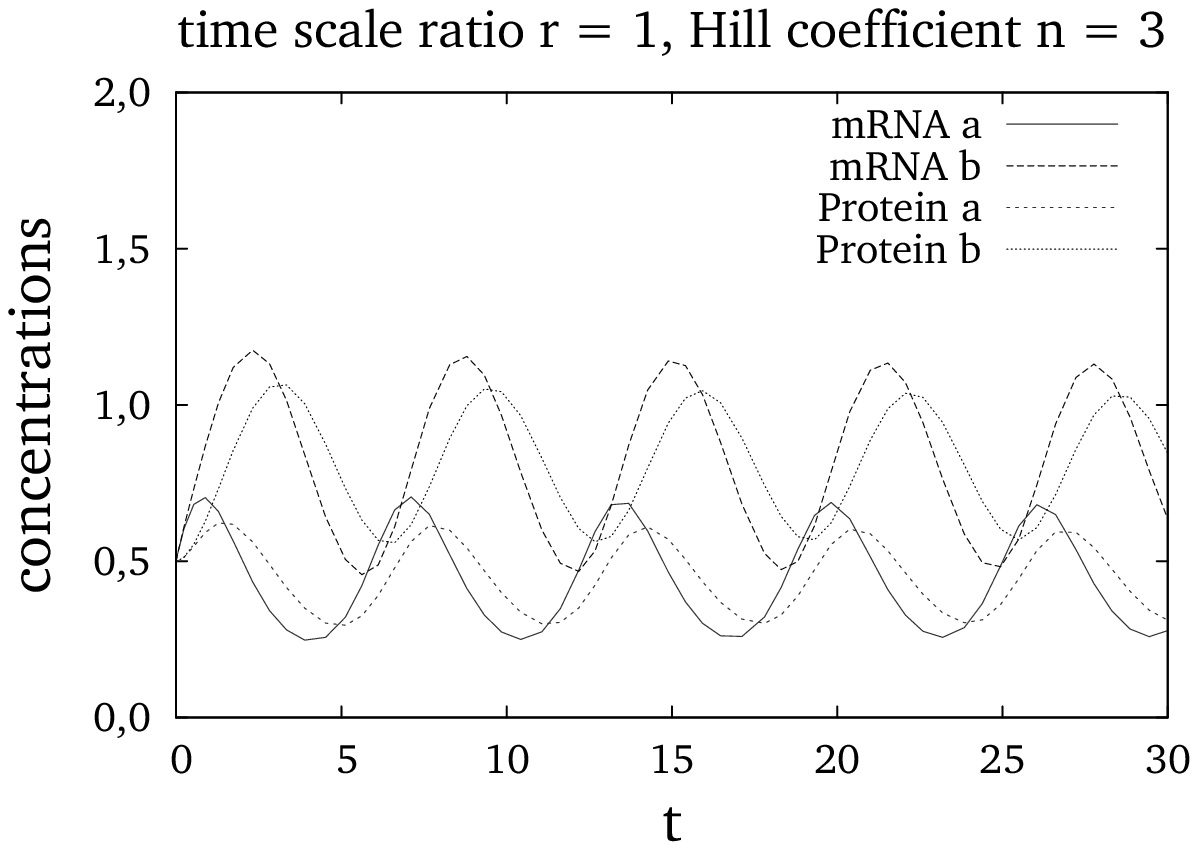}}
\subfigure{\includegraphics[width=0.23\textwidth]{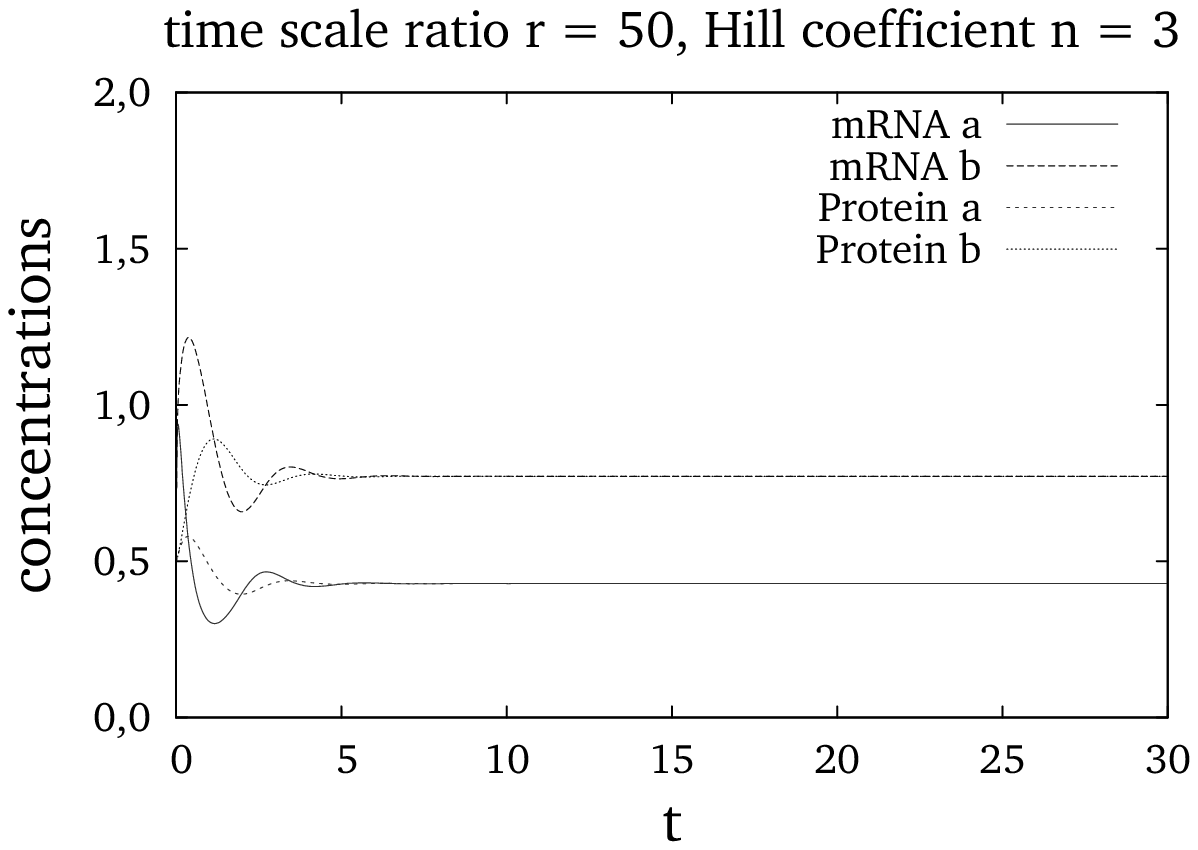}}\\
\subfigure{\includegraphics[width=0.23\textwidth]{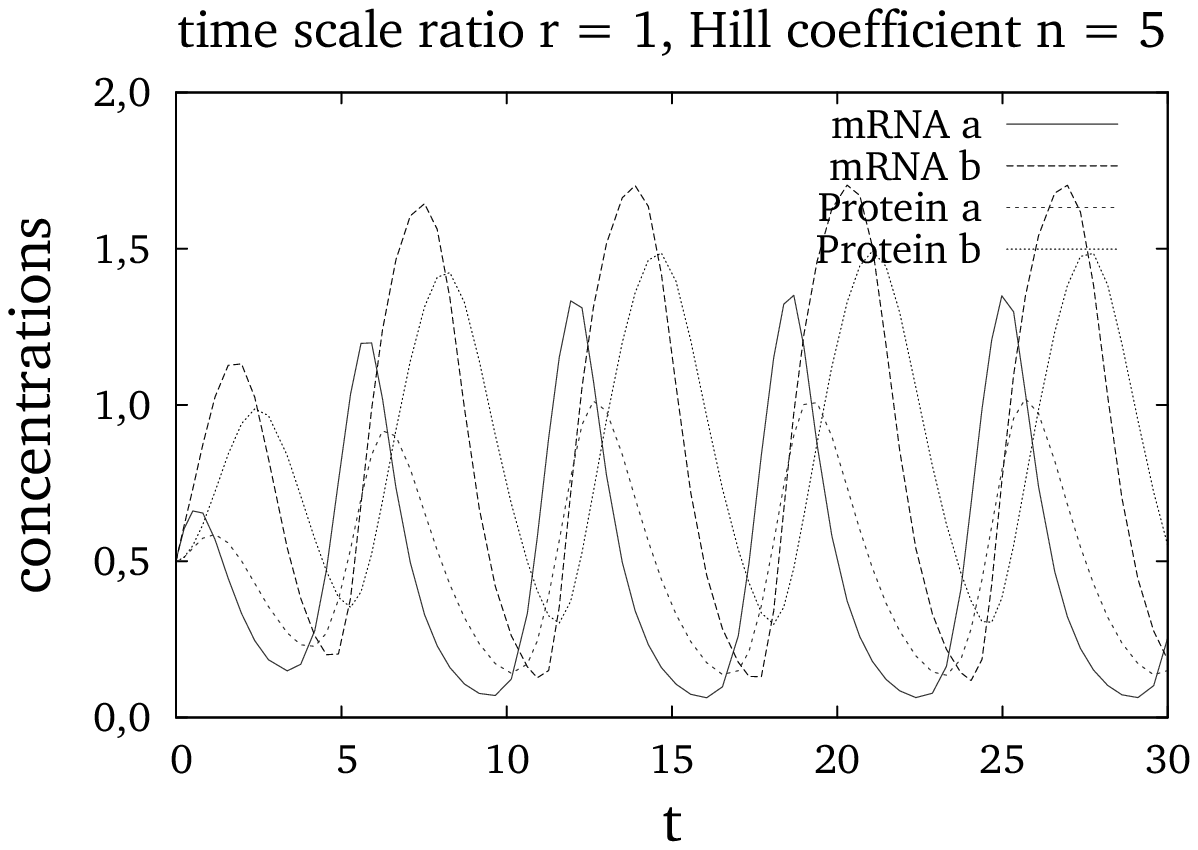}}
\subfigure{\includegraphics[width=0.23\textwidth]{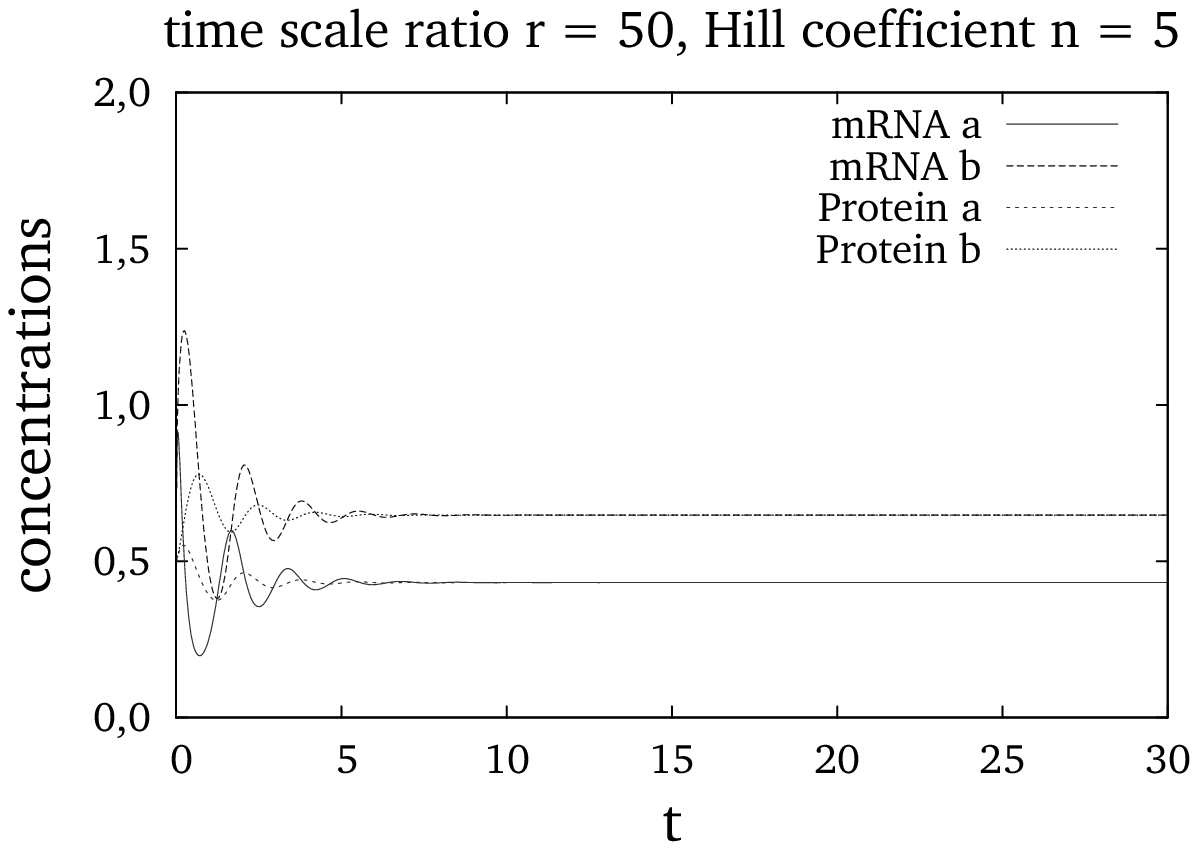}}\\
\subfigure{\includegraphics[width=0.23\textwidth]{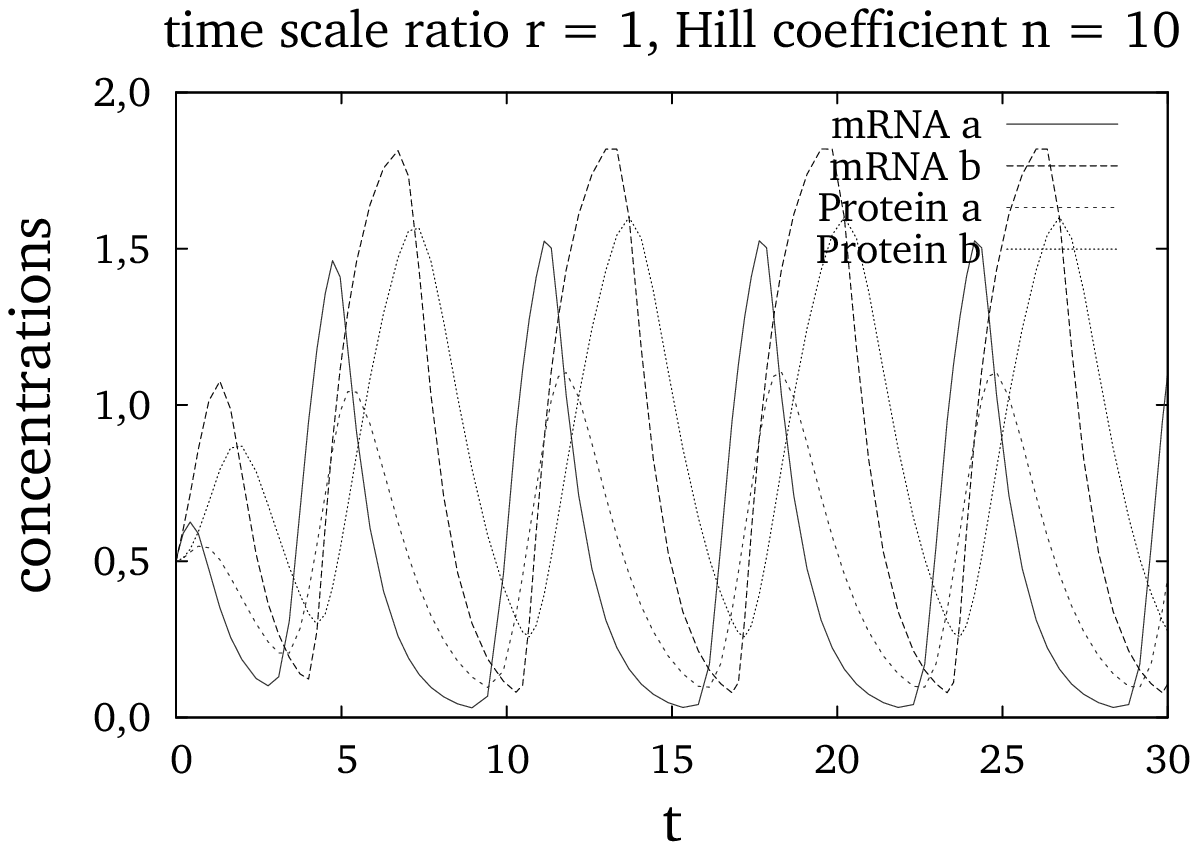}}
\subfigure{\includegraphics[width=0.23\textwidth]{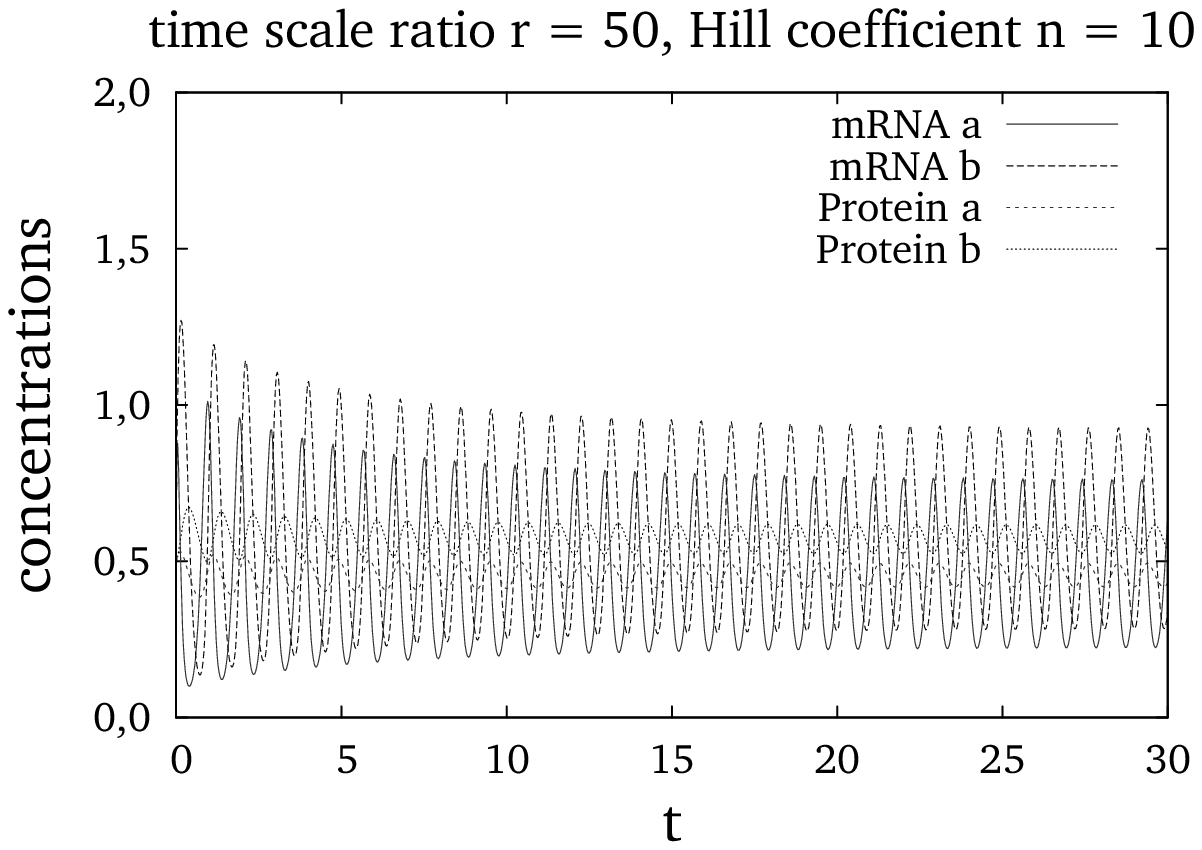}}
\caption{mRNA and protein concentrations of the network of one activator and one inhibitor for different values of $n_i$ and $r$ (equations~(\ref{ODEs}) with $F_1(P_a,P_b)=F^+(P_a)$ and $F_2(P_a)=F^-(P_a)$). The plots in the first column correspond to $r=1$, the plots in the second column to $r=50$, thus mRNA dynamics are 50 times faster, approximating the mRNA steady-state assumption. The different rows stand for different Hill coefficients: $n_i = 2,3,5,10$. With incresing $n_i$, the amplitude of the oscillation increases and stabilizes. As the separation of time scales between mRNA and proteins becomes large ($r=50$), the oscillation's frequency increases, but the amplitude of the oscillations gets smaller ($n_i = 10$) or even vanishes. (Parameter values: $m_i=2.0$, $k_i=0.5$, $\gamma_i=\delta_i=\omega_i=1.0$)}
\label{simu}
\end{figure}

Figure~\ref{varied_r_2D} shows the regions in generalized parameter space where a fixed point is unstable against oscillations. The Hopf bifurcations, which occur at the boundary between the white and the gray area, can be determined directly from the characteristic polynomial 
of the Jacobian, which gives rise to only two eigenvalues. Oscillations can only occur when the exponent parameters $\tilde{f_1}p_b$ and $\tilde{f_2}p_a$ have different signs, i.e., for the activator-inhibitor system.  Furthermore, the product of the Hill coefficients $n_a$ and $n_b$ must be large enough.  For larger time-scale ratio $r$ between mRNA and protein dynamics, the exponent parameters must be larger to obtain a Hopf bifurcation. For $r=50$, where the mRNA dynamics is quasi always in a steady-state, the activator-inhibitor network shows no oscillatory dynamics for exponent parameters $\tilde{f_1}p_b$, $\tilde{f_2}p_a \in \{1,..,4\}$, corresponding to cooperativity coefficients $n_i \in \{1,..,4\}$. Even though the generalized method uses normalised variables, the results can easily be compared to numerical simulations. Figure~\ref{simu} shows mRNA and protein concentrations for different values of $n_i$ (rows) with $r=1$ (first column) and $r=50$ (second column). As the separation of time scales between mRNA and proteins becomes larger, the oscillation frequency increases, but the amplitude of the oscillations decreases ($n_i=10$) or even vanishes ($n_i<10$).

In contrast to the Boolean model, where the state of a gene jumps instantaneously from ``off'' to ``on'' and back, the concentrations in the continuous model always change smoothly, even in the limit $n\rightarrow \infty $, where the functions $F_1$ and $F_2$ become step functions.  For this reason, the oscillation that occurs in the Boolean activator-activator (or inhibitor-inhibitor) system does not 
occur in the continuous model. Only the two fixed points that are also present in the Boolean model occur in the continuous model. For the 
same reason, the oscillation of the Boolean activator-inhibitor system can occur in the continuous model only when $F_1$ and $F_2$ are steep 
enough in order to drive the concentrations sufficiently fast through the intermediate values. Otherwise the system settles at the (only) fixed point, which is found at intermediate concentration values. 

\subsection{Model with three connections}

\begin{figure}[!ht]
\subfigure[$r = 1$]{\includegraphics[width=0.2\textwidth]{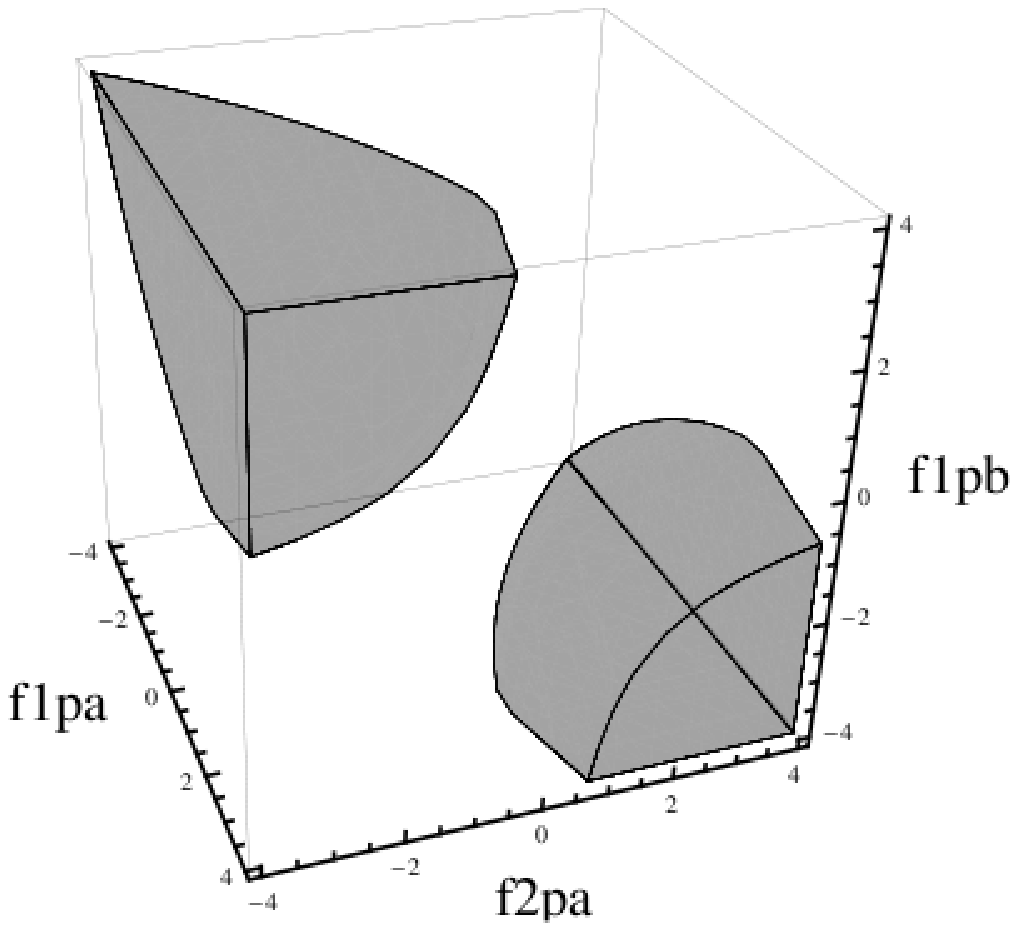}}
\subfigure[$r = 10$]{\includegraphics[width=0.2\textwidth]{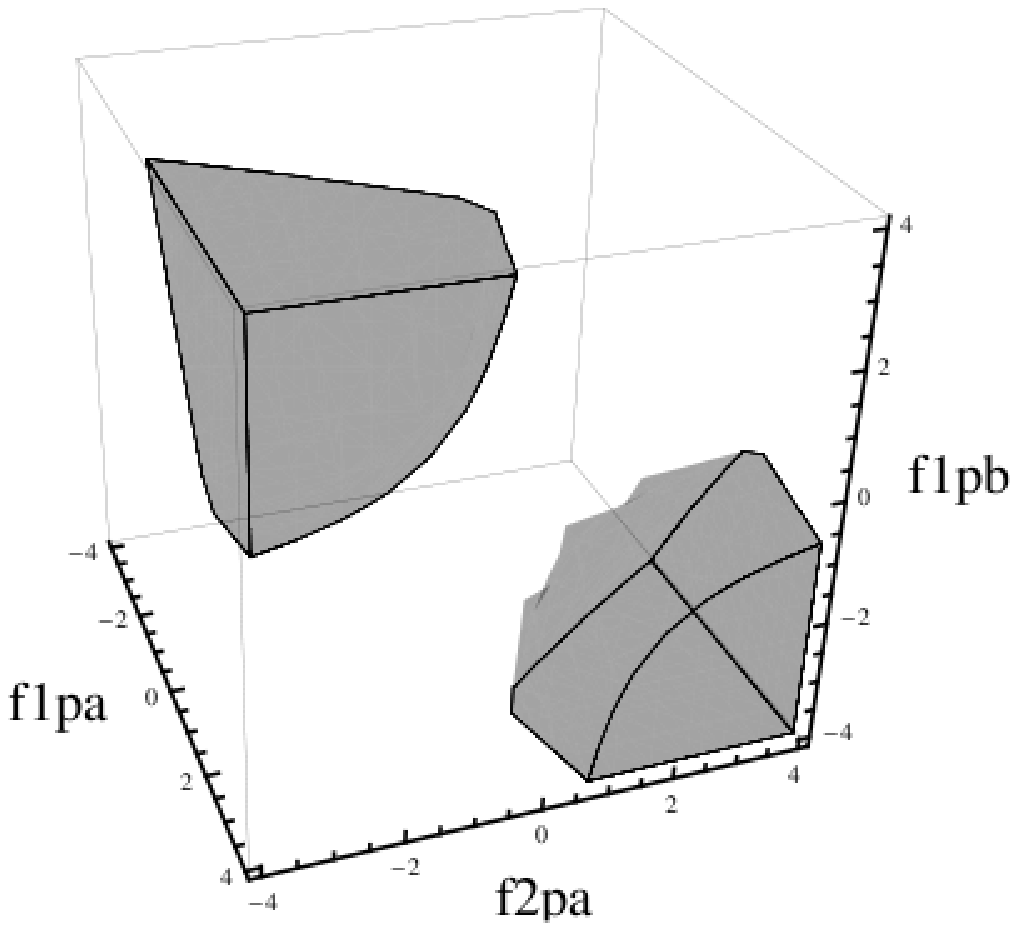}}
\subfigure[$r = 50$]{\includegraphics[width=0.2\textwidth]{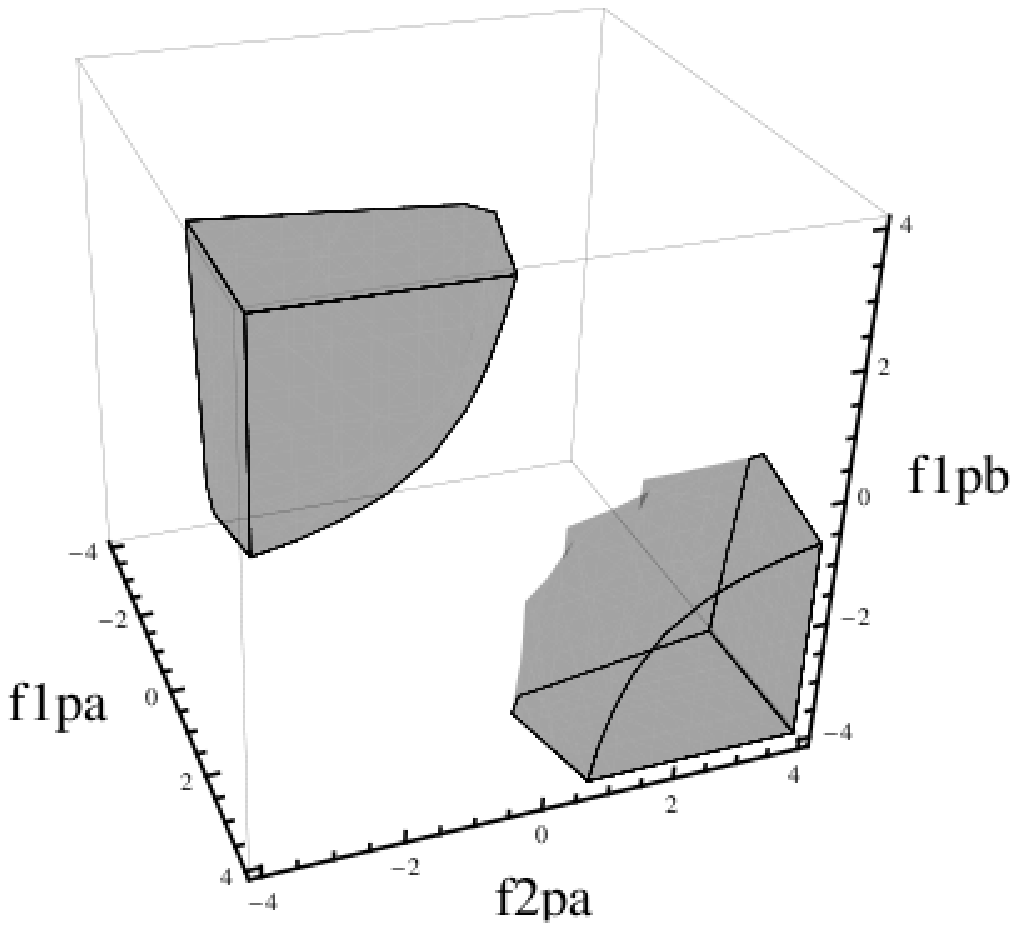}}
\caption{Regions in parameter space where the fixed point has a complex pair of unstable eigenvalues, with the other two eigenvalues being stable. Time scale ratio $r$ between mRNA and protein dynamics increases in the sequence of graphs.}
\label{varied_r}
\end{figure}

\begin{figure}[!ht]
\centering{
\includegraphics[width=0.25\textwidth]{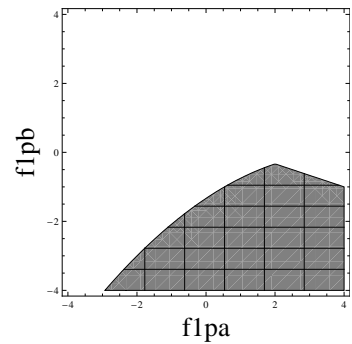}
\caption{Cross section through Fig.~\ref{varied_r} (a) at $\tilde{f_2}p_a = 3.0$.}
\label{ebene}
}
\end{figure}

Figure~\ref{varied_r} shows the generalization to the case $\tilde{f_1}p_a \neq 0$ of Fig.~\ref{varied_r_2D}. The Hopf bifurcations were calculated using the method of resultants~\cite{gross2004}. Gray areas show regions in parameter space where the fixed point has a complex pair of unstable eigenvalues, with the other two eigenvalues being stable. Figure~\ref{ebene} shows a  cross section at $\tilde{f_2}p_a= 3.0$ for $r=1$.

One sees again that a faster mRNA dynamics leads to a smaller oscillatory region in parameter space. Just as for the activator-inhibitor network, $\tilde{f_1}p_b$ and $\tilde{f_2}p_a$ must have different signs to obtain a Hopf bifurcation, corresponding to opposed regulatory functions. The additional self-input of gene~$a$ can be activating or repressing for $r=1$, but must be activating for larger ratio of time scales (e.g. $r=50$) to obtain oscillations. Larger values of the exponent parameters, and therefore larger Hill coefficients $n_i$, can make up for a large ratio of time scales between mRNA and protein dynamics, as in the system without additional self input. However, biologically realistic values of $n_i$ are in the range $n_i = 1-4$~\cite{alon}.

In order to compare the continuous model with the Boolean model, we now specify the functions $f_1$ and $f_2$ to be HillCubes. Out of the 16 possible Boolean functions of two variables for $F_1(P_a,P_b)$, only 10 actually depend on the values of both variables. We discuss in the 
following these 10 functions and compare the Boolean dynamics with the dynamics due to HillCubes. We restrict ourselves to the case that $a$ 
activates $b$, i.e., $\tilde{f_2}p_a > 0$. The case that $a$ inhibits $b$ can always be mapped on the first case in a Boolean model by inverting the states of nodes and changing the sign of the appropriate connections.

The Boolean functions $a$~AND~$b$ and $a$~OR~$b$ give rise to the fixed points 00 and 11 in the Boolean model. Using HillCubes, we obtain $\tilde{f_1}p_b > 0$ for both functions, which means that the continuous model cannot have a Hopf bifurcation. Whether the continuous model has one or two stable fixed points, depends on the parameter values. 

The Boolean functions NOT~$a$~AND~$b$ and NOT~$a$~OR~$b$ give rise to one fixed point (00 and 11, respectively) and one cycle involving the 
states 01 and 01 in the Boolean model.  Using HillCubes, we obtain $\tilde{f_1}p_b > 0$ for both functions, which means that the continuous model cannot have a Hopf bifurcation. This situation is analogous to the activator-activator loop, where the cycle present in the Boolean model does not occur in the continuous model. 

\begin{figure}[!ht]
\centering
\subfigure[ $P_a$~ANDNOT~$P_b$]{\includegraphics[width=0.23\textwidth]{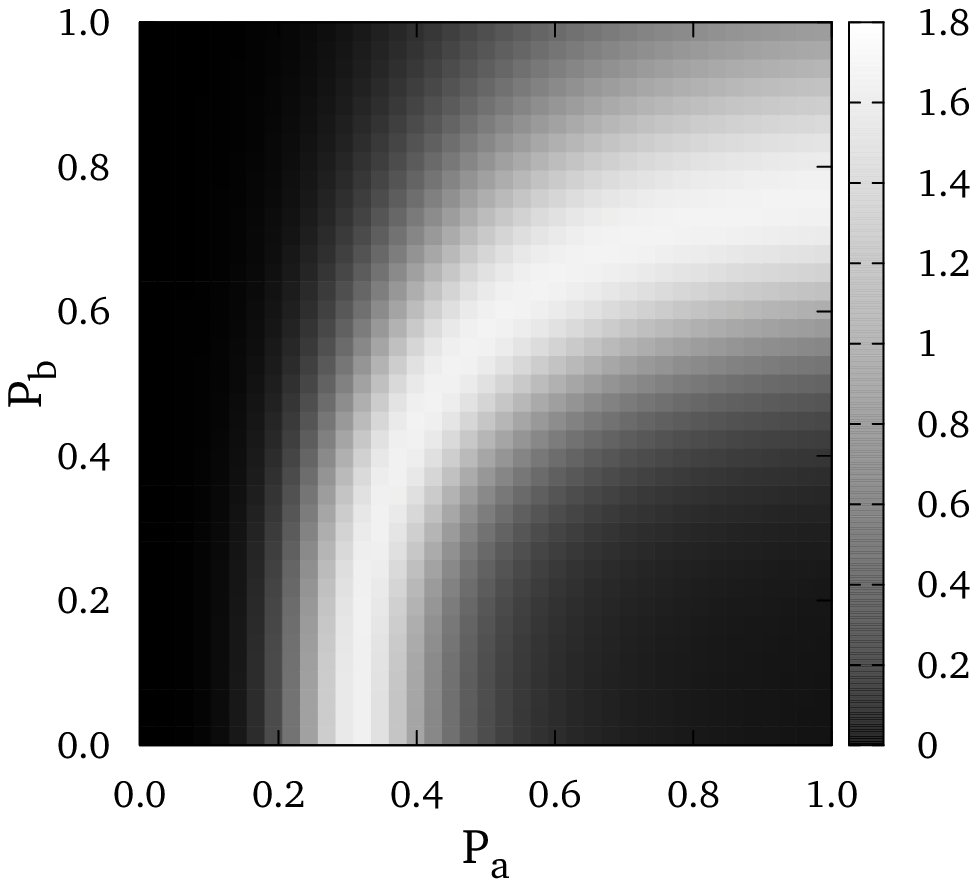}}
\subfigure[ $P_a$~NOR~$P_b$]{\includegraphics[width=0.23\textwidth]{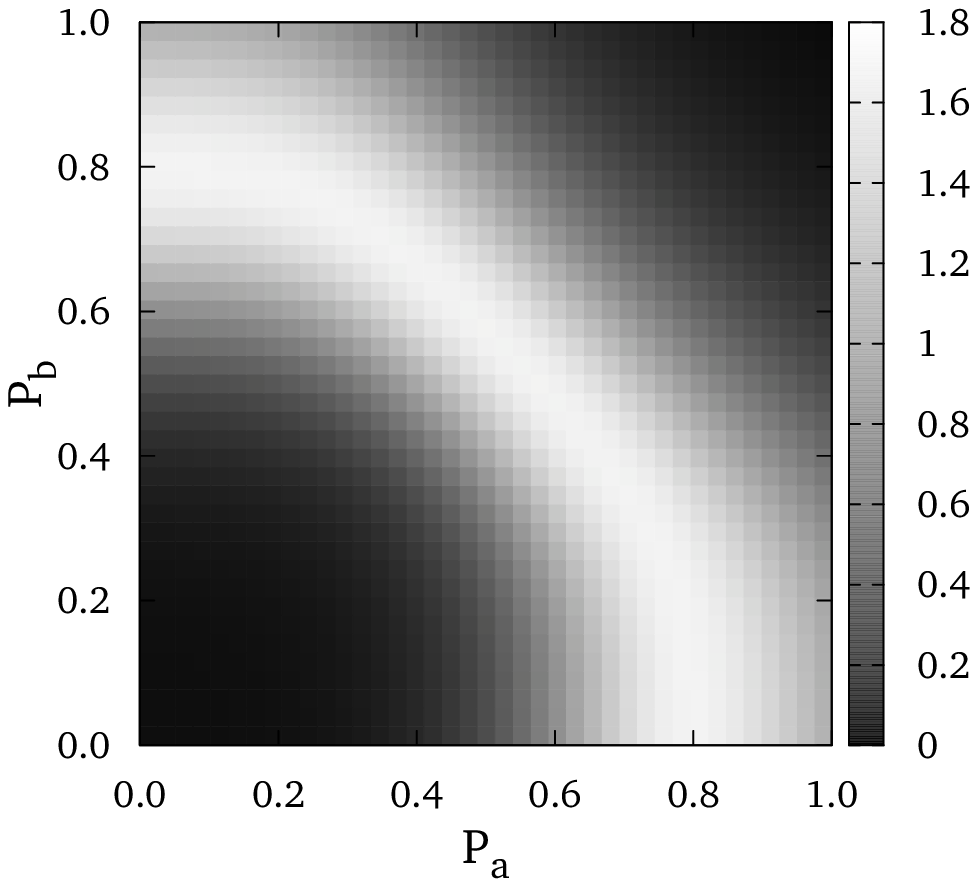}}
\subfigure[ $P_a$~XOR~$P_b$]{\includegraphics[width=0.23\textwidth]{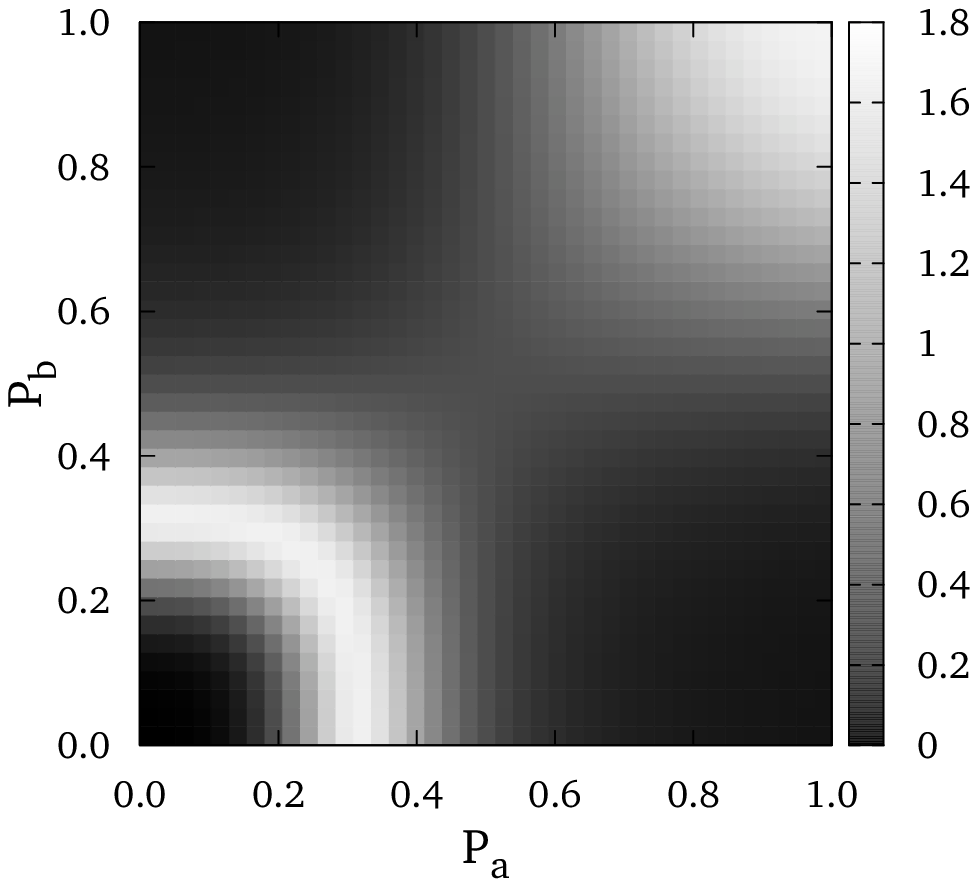}}
\caption{Two-dimensional input functions $F=P_a~\mbox{AND NOT}~P_b$, $F=P_a~\mbox{NOR}~P_b$ and $F=P_a~\mbox{XOR}~P_b$ for $k_i= 0.5$ and $n_i = 3$, projected to the two-dimensional surface. The light-colored areas mark those regions in parameter space where $F_2(P_a)$ is steep, implying intermediate values of the mRNA $a$ concentration.}
\label{hc_F2}
\end{figure}

The Boolean functions $a$~AND~NOT~$b$ and $a$~OR~NOT~$b$ give rise to one fixed point (00 and 11, respectively), which is a global attractor, in the Boolean model. Using HillCubles, we obtain $\tilde{f_1}p_b < 0$ and $\tilde{f_1}p_a > 0$. The signs of the exponent parameters are such that a Hopf bifurcation is possible. When the concentration of transcribed mRNA is projected onto a two-dimensional surface (Fig.~\ref{hillcubes}), one can see that the two functions $a$~AND~NOT~$b$ and $a$~OR~NOT~$b$ produce significant mRNA concentrations in the shaded area of Fig.~\ref{ebene}. An optimal condition for a Hopf bifurcation are intermediate values of the mRNA 
concentration, since the function $F_2$ is steep in this region. Figure~\ref{hc_F2} shows regions in parameter space, where  $F_2(P_a)$ is steep, for the function $a$~AND~NOT~$b$ (and for two other functions to be discussed below).  

\begin{figure}[!ht]
\centering{
\subfigure[ $P_a$~ANDNOT~$P_b$]{\includegraphics[width=5cm]{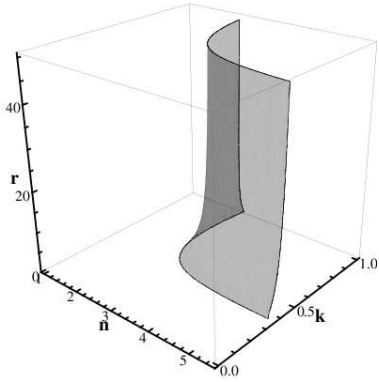}}
\subfigure[ $P_a$~NOR~$P_b$]{\includegraphics[width=5cm]{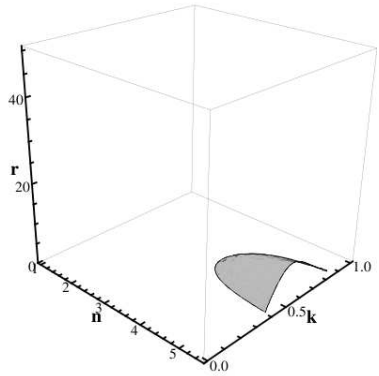}}
\caption{Hopf bifurcation surface of the two-dimensional input functions $F~=~a$~ANDNOT~$b$ and $F~=~a$~NOR~$b$. Within the surfaces, the fixed point has a complex pair of unstable eigenvalues, with the other two eigenvalues being stable. Using generalized functions $\tilde f$ of $F~=~a$~ANDNOT~$b$ at a fixed protein concentration $P_a^* = P_b^* = 0.5$ and $F~=~a$~NOR~$b$ at $P_a^* = 0.4, P_b^* = 0.6$, we varied $n_i$, $k_i$ and $r$. The choice of $P_a^*$ and $P_b^*$ is done with the help of Fig.~\ref{hc_F2}, to be in an region in parameter space which is likely to induce oscillations. The input function $F~=~a$~XOR~$b$ does not provide a Hopf bifurcation.}
\label{nkr}
}
\end{figure}

\begin{figure}[!ht]
\centering
\includegraphics[width=0.23\textwidth]{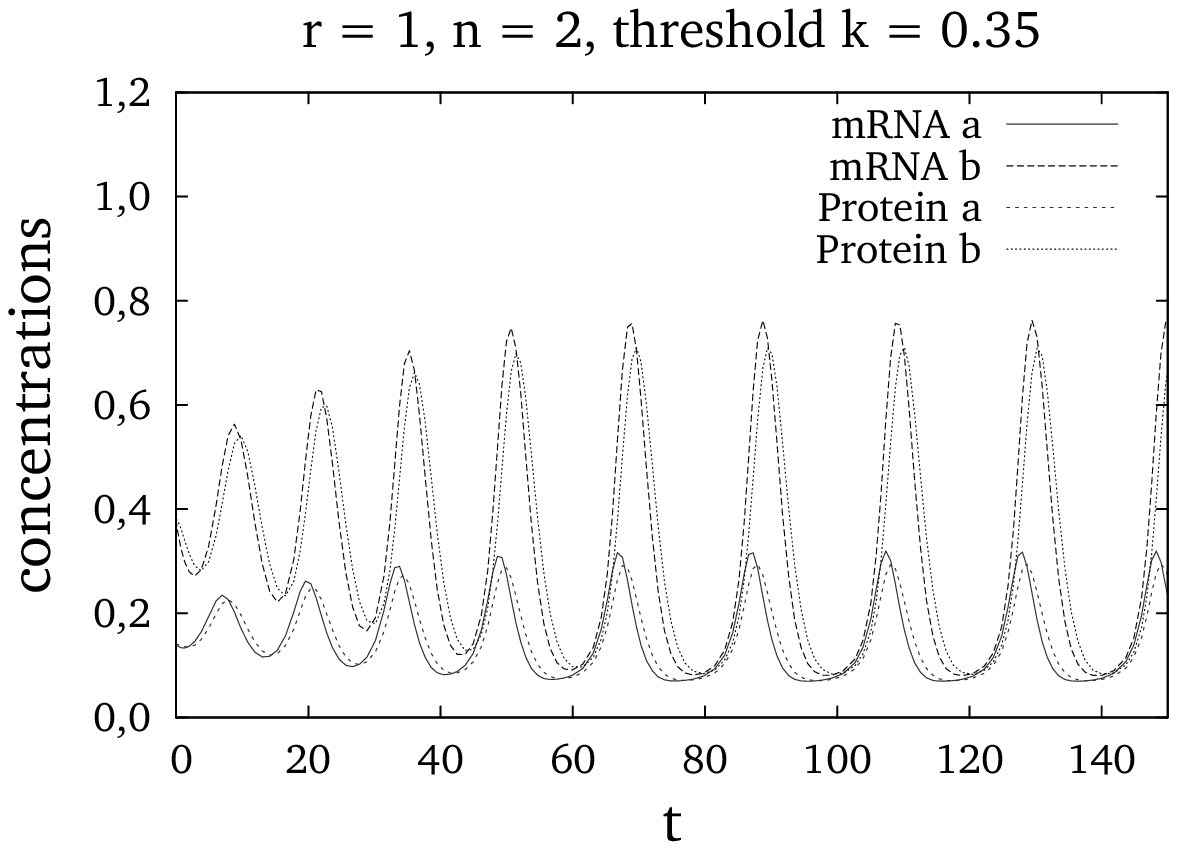}
\includegraphics[width=0.23\textwidth]{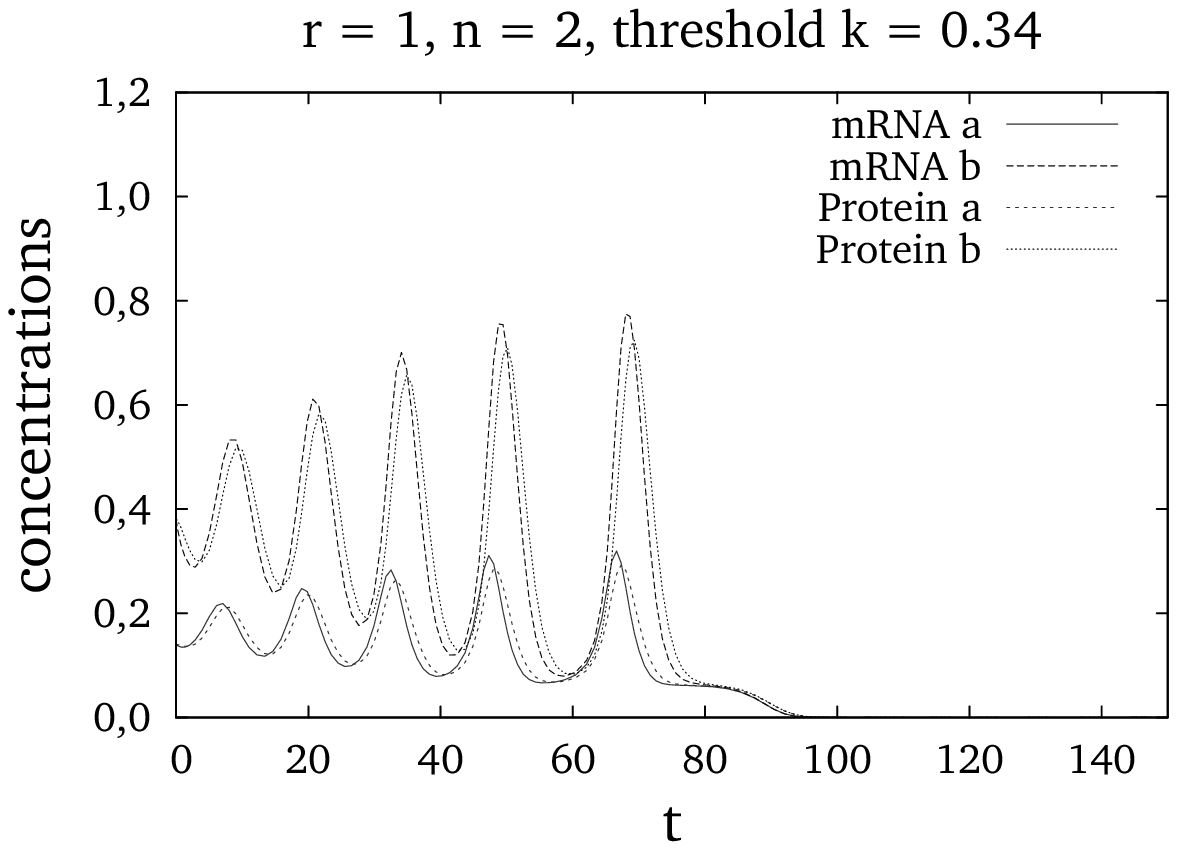}
\caption{mRNA and protein concentrations of the two-gene module with additional self-input and $F_1=P_a~\mbox{AND NOT}~P_b$  (equations~(\ref{ODEs})) for $n_i=2$ and $r=1$. For $k_i = 0.35$ exists a complex pair of unstable eigenvalues, and we obtain sustained oscillations. At $k_i = 0.34$, the oscillation has become unstable, and the periodic orbit grows until it collides with the fixed point $0$. (Parameter values: $m_i=2.0$,   $\gamma_i=\delta_i=\omega_i=1.0$) }
\label{simu_ANDNOT}
\end{figure}

In order to determine the parameter regions that place the system in the shaded area of Fig.~\ref{ebene}, we varied the parameters $n_i$, 
$k_i$ and $r$ at fixed protein concentrations $P_a^*$ and $P_b^*$. The values of $P_a^*$ and $P_b^*$ were chosen such that the system is 
placed in the light-colored areas of Fig.~\ref{hc_F2}, where oscillations are most likely. The result is shown in Fig.~\ref{nkr}, and  numerical simulations are shown in Fig.\ref{simu_ANDNOT}: There exists a complex pair of unstable eigenvalues for $n_i=2$ and $r=1$. For $k_i = 0.35$ we obtain sustained oscillations. However, for this set of parameter values, the system is close to a bifurcation: at $k_i = 0.34$ the periodic orbit grows until it collides with the fixed point~$0$.

\begin{figure}[!ht]
\centering
\subfigure{\includegraphics[width=0.23\textwidth]{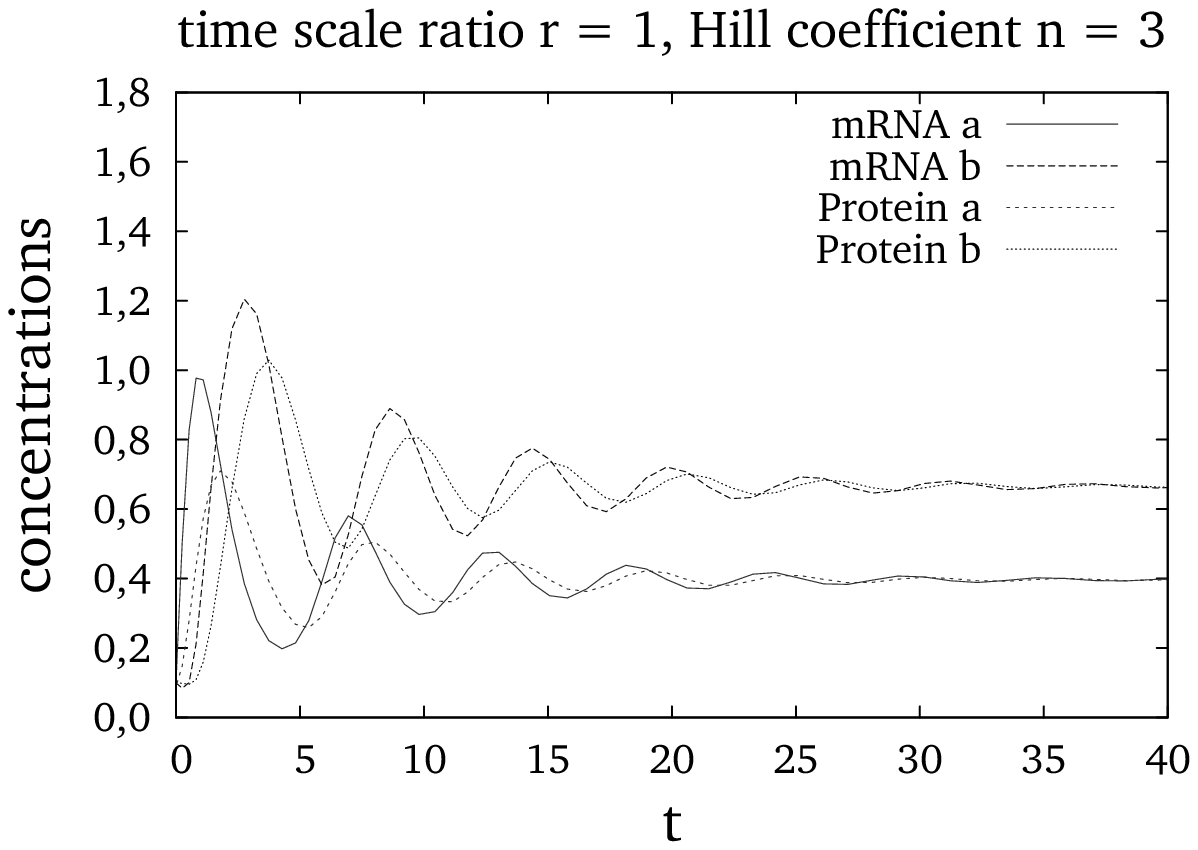}}
\subfigure{\includegraphics[width=0.23\textwidth]{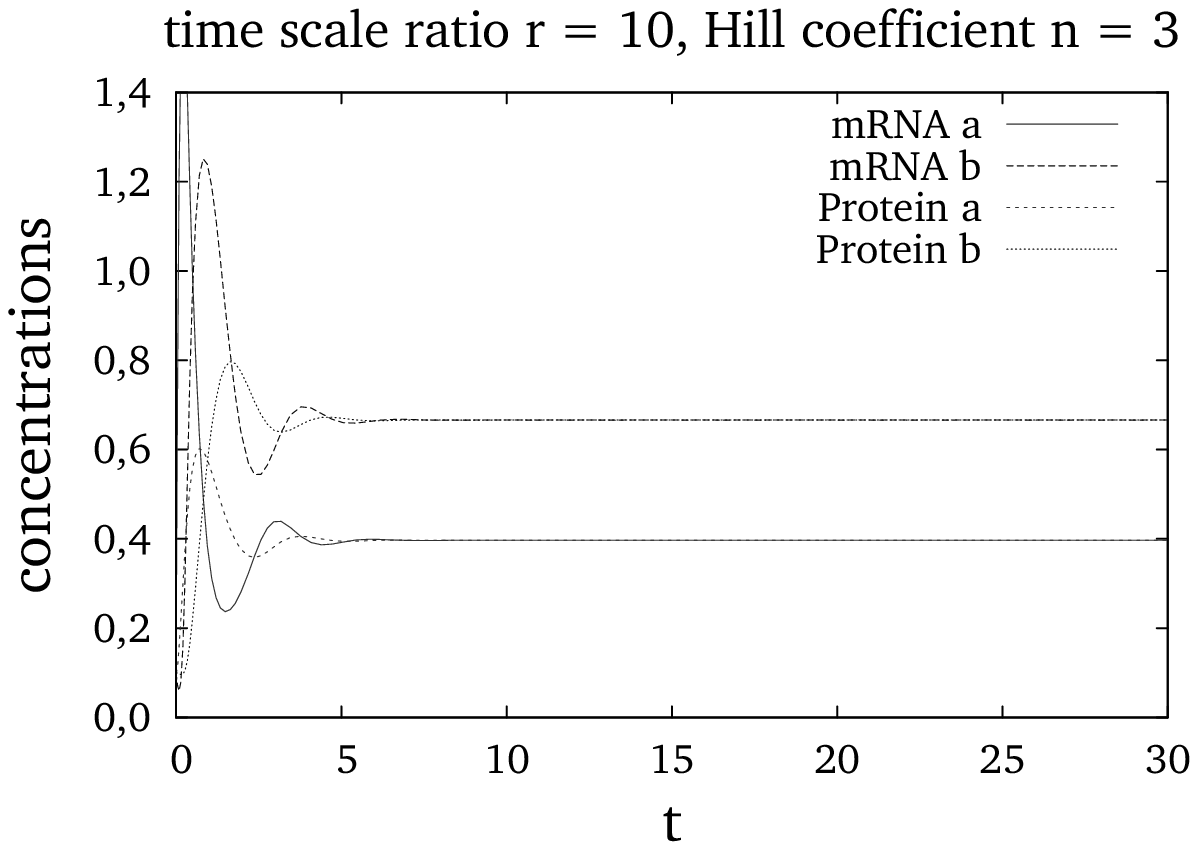}}\\
\subfigure{\includegraphics[width=0.23\textwidth]{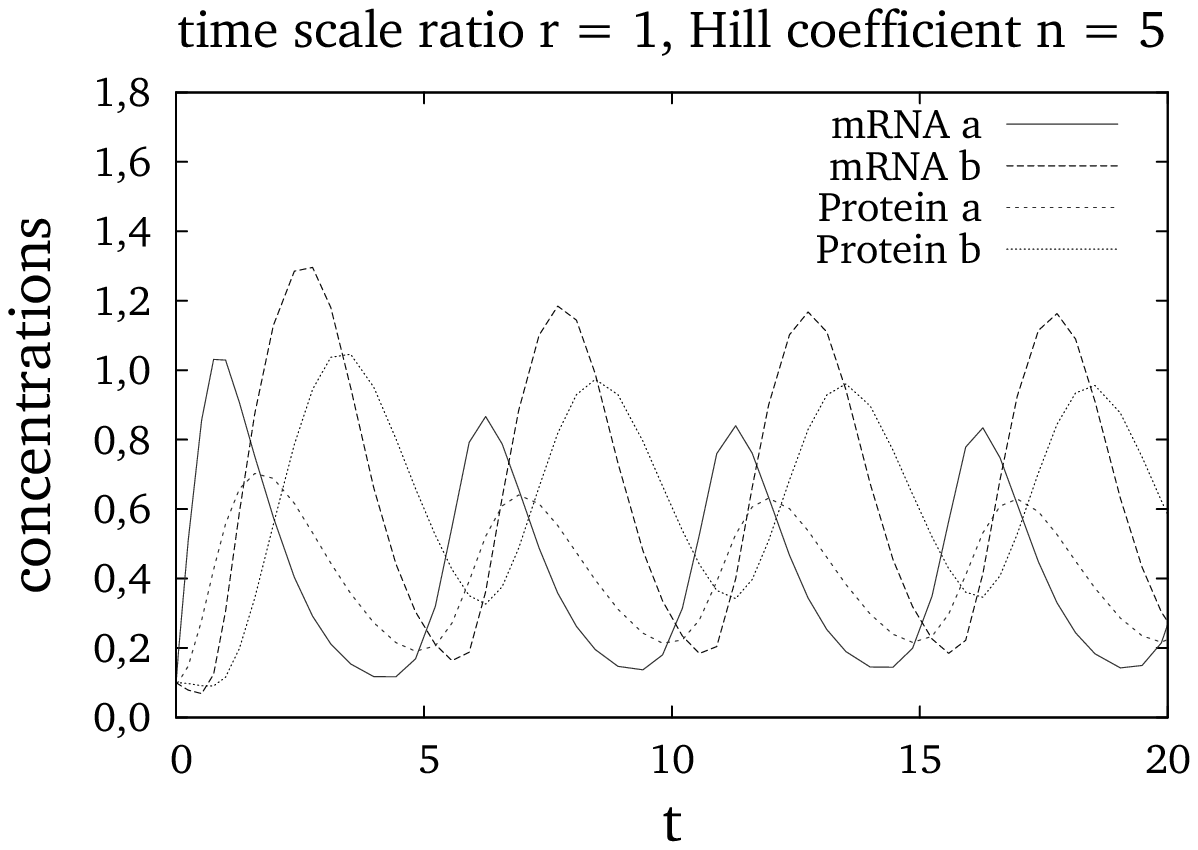}}
\subfigure{\includegraphics[width=0.23\textwidth]{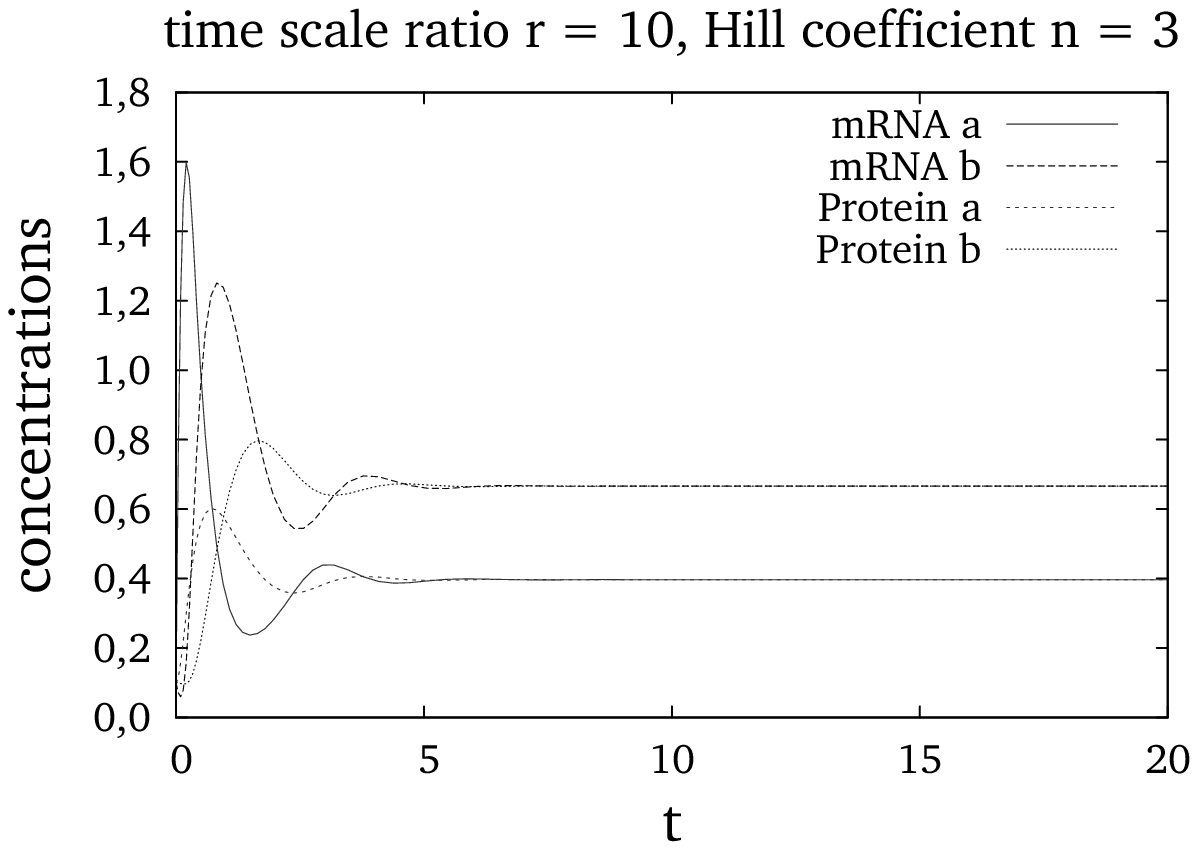}}\\
\subfigure{\includegraphics[width=0.23\textwidth]{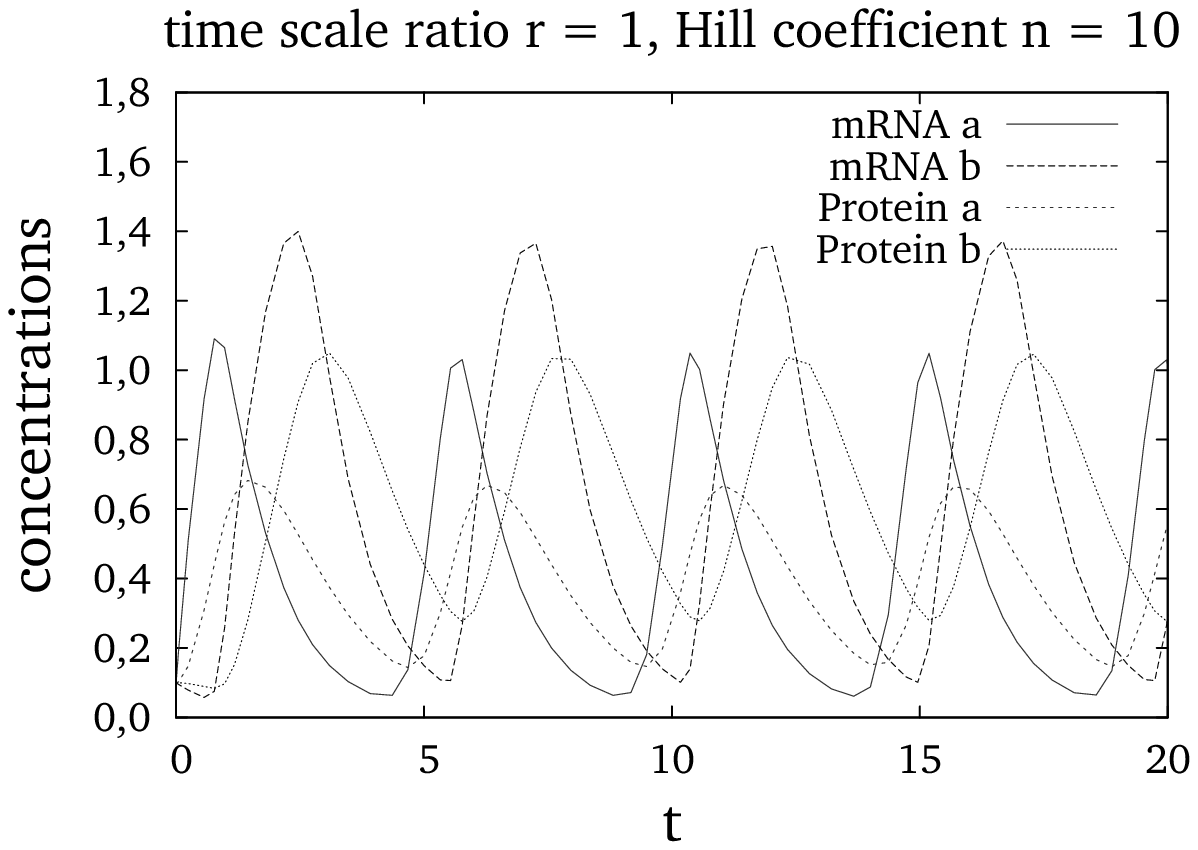}}
\subfigure{\includegraphics[width=0.23\textwidth]{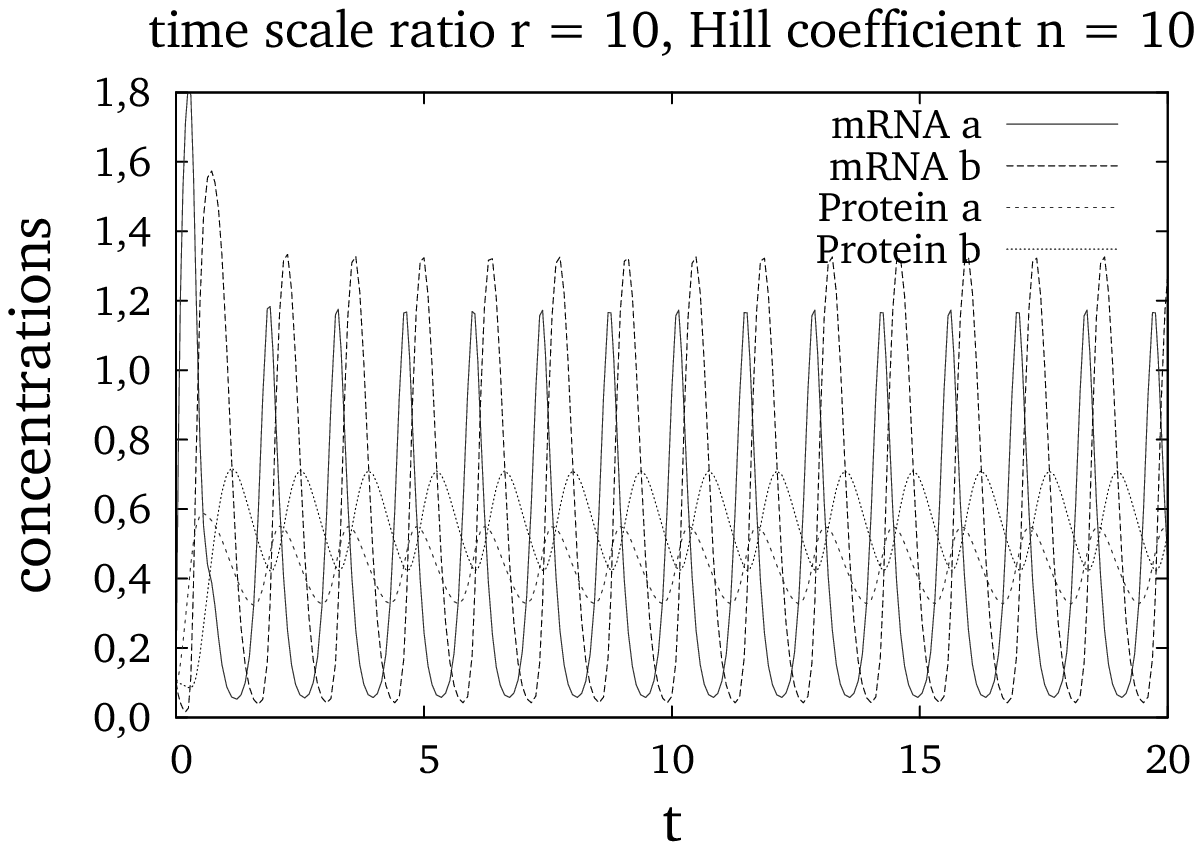}}\\
\caption{mRNA and protein concentrations of the two-gene module with additional self-input and $F_1=P_a~\mbox{NOR}~P_b$  (equations~(\ref{ODEs})) for different values of $n_i$ and $r$. The plots in the first column correspond to $r=1$, the plots in the second column to $r=10$, thus mRNA dynamics are 10 times faster, corresponding to Fig.\ref{varied_r}(b). The different rows stand for different Hill coefficients: $n_i = 2,3,5$. With incresing $n_i$, the amplitude of the oscillation increases and stabilizes. As the separation of time scales between mRNA and proteins becomes larger ($r=10$), the oscillations vanish for small values of $n_i$. (Parameter values: $m_i=2.0$, $k_i=0.5$, $\gamma_i=\delta_i=\omega_i=1.0$) }
\label{simu_NOR}
\end{figure}

The Boolean functions $a$~NOR~$b$ and $a$~NAND~$b$ give rise to a cycle, which is a global attractor, in the Boolean model. Using HillCubles, we obtain $\tilde{f_1}p_b < 0$ and $\tilde{f_1}p_a <0$. The signs of the exponent parameters are such that a Hopf bifurcation is possible.  The projections of the functions on a two-dimensional surface and the parameter regions that place the system in the shaded area of Fig.~\ref{ebene} are again shown in Figs~\ref{hc_F2} and \ref{nkr}. The results are compared to numerical simulations, shown in Fig.~\ref{simu_NOR}(b). In contrast to the previous case, the parameter region where oscillations are possible is much smaller. Oscillations are only possible when mRNA dynamics is sufficiently slow and when the Hill coefficient $n_b$ is above~$2$. 

Finally, we consider the Boolean functions $a$~XOR~$b$ and $a$~XNOR~$b$. In contrast to the other functions discussed so far, these functions are not canalyzing but change their output value whenever one of the input values is changed. In the Boolean model, these functions give rise to one fixed point and to one cycle that comprises the three remaining states. The signs of $\tilde{f_1}p_b $ and $\tilde{f_1}p_a$ in the continuous model can now be positive or negative, depending on the parameter values. An oscillation is thus possible in principle. Based on Fig.~\ref{hc_F2} we chose fixed point concentrations likely to induce oscillations to calculate the surface of the Hopf bifurcation. But in contrast to the two sets of functions discussed in the previous section, we did not find an oscillatory region. 

\section{Discussion and conclusions}
\label{discussion}

By using the method of generalized models to study the dynamics of simple two-gene regulatory network components, we could establish general conditions for the occurrence of Hopf bifurcations, which give rise to oscillatory dynamics. Apart from the signs of the regulatory interactions, the only relevant parameters in this general description are the Hill coefficient and the time scale ratio between mRNA and protein dynamics. By comparing the different types of interactions to two-node Boolean models, we found that the occurrence of a cycle in the Boolean model is neither necessary nor sufficient for the occurrence of an oscillation in the continuous model. 

Our results combine and generalize the findings of several previous studies of such systems. The studies by Widder~et~al.~\cite{widder2007} and Polynikis~et~al.~\cite{poly2009} were focused on a two-gene system with only two connections, and they found that oscillations can only occur in the activator-inhibitor case, and only for large Hill coefficients $n \geq3$~\cite{widder2007}, or more precisely for $n_a \cdot n_b > 4$~\cite{poly2009}. (Choice of other parameters: $\gamma_i = \delta_i = \omega_i = 1.0$.) Polynikis~et~al.~\cite{poly2009} found that the system can be driven through the Hopf bifurcation by varying the time scales of the mRNA and the proteins.  Widder~et~al.~\cite{widder2007} emphasize that the domain in parameter space that contains a limit cycle becomes larger with increasing cooperativity as expressed by higher Hill coefficients. All these findings are contained concisely in our Fig.~\ref{varied_r_2D}.

The two-gene system with three connections was studied by Del~Vecchio~\cite{vecchio2007}, for the case $\tilde{f_1}p_b < 0$, $\tilde{f_1}p_a > 0$, $\tilde{f_2}p_a > 0$, by using generalized Hill functions. The extensive bifurcation analysis shows that oscillations occur over a larger parameter range when the mRNA dynamics is slower. This result is a special case of our very general result shown in Fig.~\ref{varied_r}.

Norrell~et~al.~\cite{norrell2007}, studied simple rings of four genes, and rings where one node has an additional self-input.  They included a time delay to model translation, instead of including additional equations for the mRNA concentrations. They found that the time delay must be sufficiently large for oscillations to occur, which agrees with our finding that $r$ must be sufficiently small. Furthermore, they found that continuous systems can exhibit stable oscillations in cases where Boolean reasoning would suggest otherwise, and that on a ring periodic cycles of Boolean systems do not exist in continuous systems when the oscillation is not stable against fluctuations in the update time. This finding is a generalization of the result described above that the two-gene activator-activator system has no oscillations. In a different publication~\cite{norrell2009}, they point out the importance of the detailed form of the continuous functions for obtaining nontrivial dynamical patterns. This finding is related to our finding that the Hill coefficents $n_i$ must be sufficiently large for oscillations to occur. 

Mochizuki~\cite{mochizuki2005} investigated random networks with larger numbers of nodes. He found that the number of different steady states increases with the number of self-regulatory genes. He furthermore found that many of the periodic oscillations observed in the Boolean network are not present in the continuous model of gene regulatory networks and therefore the predictions of Boolean models can become unrealistic or too complex for larger networks when compared to those of the corresponding ODE models. The latter finding is conform with the findings for the simple two-gene model. 

When all these investigations are taken together, there appears to be no simple criterion for deciding whether a periodic dynamical behavior in a Boolean model has an equivalent in a continuous model. In the examples investigated in this paper, there are two cases where the periodic dynamics of the Boolean model is ``reliable'' in the sense that fluctuations in the update times of the two nodes do not destroy the dynamical cycle. The first case is the activator-inhibitor system, the second is  the NOR and NAND function. The corresponding continuous model has periodic oscillations whenenver the Hill coefficients are large enough and the time scale of the mRNA is slow enough. However, a reliable oscillation in the Boolean model is not a necessary requirement for an oscillation in the continuous model, as shown for 
the XOR and XNOR functions (where the oscillation in the Boolean model is not reliable), and for the AND NOT and OR NOT functions, where the Boolean model has a global fixed point.  

Certainly,  more research is needed to establish more general conditions under which larger networks with continuous dynamics are conform with their Boolean counterpart.


\end{document}